%% file: paper.tex
\documentclass[prodmode,acmtoms]{acmsmall}

\acmVolume{X}
\acmNumber{Y}
\acmArticle{00}
\acmYear{2015}
\acmMonth{10}

\usepackage{amsmath}
\usepackage{url}
\usepackage{color}
\usepackage[ruled,linesnumbered]{algorithm2e}
\SetAlFnt{\small}
\SetAlCapFnt{\small}
\SetAlCapNameFnt{\small}
\SetAlCapHSkip{0pt}
\IncMargin{-\parindent}

\SetEndCharOfAlgoLine{\ }

\usepackage[caption=false]{subfig}

\title{Pipelined Iterative Solvers with Kernel Fusion \\for Graphics Processing Units}

\author{KARL RUPP,
JOSEF WEINBUB,
ANSGAR J\"UNGEL,
and TIBOR GRASSER
\affil{\\TU Wien}
}

\usepackage{listings}
\usepackage{color}

\definecolor{mygray}{rgb}{0.5,0.5,0.5}

\begin{document}

\begin{abstract}
We revisit the implementation of iterative solvers on discrete graphics processing units and demonstrate the benefit of implementations using extensive kernel fusion for pipelined formulations over conventional implementations of classical formulations.
The proposed implementations with both CUDA and OpenCL are freely available in ViennaCL and are shown to be competitive with or even superior to other solver packages for graphics processing units.
Highest performance gains are obtained for small to medium-sized systems, while our implementations are on par with vendor-tuned implementations for very large systems.
Our results are especially beneficial for transient problems, where many small to medium-sized systems instead of a single big system need to be solved.
\end{abstract}




\acmformat{Karl Rupp, Josef Weinbub, Ansgar J\"ungel, and Tibor Grasser, 2015. Pipelined iterative solvers with kernel fusion for graphics processing units.}

\markboth{K.~Rupp, J.~Weinbub, A.~J\"ungel, T.~Grasser}{Pipelined Iterative Solvers with Kernel Fusion}

\maketitle

\begin{bottomstuff}
This work is supported by the Austrian Science Fund (FWF) through the grants P23296 and P23598.
The authors also thank AMD and NVIDIA for hardware donations and Joachim Sch\"oberl for providing access to a system equipped with two NVIDIA K20m for benchmarking purposes.

Author's addresses: Karl Rupp, Josef Weinbub and Tibor Grasser, Institute for Microelectronics, 
TU Wien,
Gu\ss hausstra\ss e 27-29/E360, A-1040 Wien, Austria;
email: \{rupp, weinbub, grasser\}@iue.tuwien.ac.at;
Ansgar J\"ungel, Institute for Analysis and Scientific Computing, 
TU Wien,
Wiedner Hauptstra\ss e 8-10/E101, A-1040 Wien, Austria;
email: juengel@asc.tuwien.ac.at
\end{bottomstuff}

\input{intro.tex}

\input{techniques.tex}

\input{solvers.tex}

\input{benchmarks.tex}

\input{conclusion.tex}

\bibliographystyle{ACM-Reference-Format-Journals}
\bibliography{ref}

\end{document}

%% file: intro.tex
\section{Introduction}

The need for the solution of a linear system of equations described by a sparse matrix $A$ and a right hand side vector $b$ is ubiquitous in computational science and engineering.
Most prominently, discretizations of linear partial differential equations by means of the finite element or the finite volume method directly lead to such systems.
Smaller-sized systems may be solved using sparse direct solvers, whereas iterative solvers are preferred or even necessary for large systems, eventually supplemented by preconditioning techniques of various degrees of sophistication.

The fine-grained parallelism of iterative solvers from the family of Krylov methods is particularly attractive for massively parallel hardware such as graphics processing units (GPUs), whereas much more effort is required to expose the parallelism in sparse direct solvers appropriately \cite{Kim:sparse-direct-GPU,Schenk:studies-gpu}.
Sparse matrix-vector products - essential parts of Krylov methods - have been studied in detail for GPUs \cite{Ashari:SpMV-gpus,baskaran:sparse-matvec-ibm,Bell:SpMV-on-throughput-oriented-processors,Greathouse:CSR-adaptive} and for INTEL's many-integrated core (MIC) architecture \cite{Liu:Efficient-SpMV,Saule:SpMV-Xeon-Phi}, based on which a unified format also well-suited for multi-core processors has been proposed recently \cite{Kreutzer:Unified-sparse-matrix-format}.
Similarly, vendor-tuned implementations of the vector operations required in addition to the sparse matrix-vector products for implementing sparse iterative solvers from the family of Krylov methods are available.
A disadvantage of current accelerators is their connection to the host system via the PCI-Express bus, which is often a bottleneck both in terms of latency as well as bandwidth.
This mandates a certain minimum system size to amortize the overhead of data transfer through the PCI-Express bus in order to obtain any performance gains over an execution on the host.

Two programming models are currently in widespread use for general purpose computations on GPUs:
CUDA is a proprietary programming model for NVIDIA GPUs~\cite{Nickolls:CUDA} providing its own compiler wrapper, whereas OpenCL is a royalty-free open standard maintained by the Khronos Group\footnote{Khronos Group, OpenCL: \url{http://www.khronos.org/opencl/}} and is typically provided as a shared library.
Although OpenCL can also be used for NVIDIA GPUs, the richer CUDA toolchain has resulted in a higher share of research on general purpose computations on GPUs using CUDA.
Also, slight performance differences of CUDA and OpenCL, caused by different degrees of compiler optimizations or differences in the implementation rather than through differences in the programming model, have been reported~\cite{Fang:performance-cuda-opencl,karimi:performance-cuda-opencl}.
Automated translators such as Swan~\cite{Harvey20111093} or CU2CL~\cite{Martinez:CU2CL} have been developed to reduce the maintenance effort of CUDA and OpenCL branches.
However, only a subset of CUDA and OpenCL is supported by these translators, limiting their applicability particularly for highly optimized implementations.
Directives-based approaches for general purpose computations on GPUs are the OpenACC\footnote{OpenACC: \url{http://www.openacc-standard.org/}} and OpenMP\footnote{OpenMP: \url{http://openmp.org/}} standards.
Broad compiler support for both standards in the context of GPUs is, however, not yet available.
Consequently, portable software libraries targeting GPUs are currently driven into providing support for both CUDA and OpenCL, for example PARALUTION\footnote{PARALUTION library: \url{http://www.paralution.com/}}, VexCL\footnote{VexCL library: \url{https://github.com/ddemidov/vexcl/}}, or ViennaCL\footnote{ViennaCL library: \url{http://viennacl.sourceforge.net/}}.

A substantial amount of research has been conducted on various preconditioning techniques for iterative solvers on GPUs including algebraic multigrid \cite{Bell:AMG,Gandham:GPU-Aggregation-AMG,Richter:GPU-AMG,Wagner:AMG}, incomplete factorizations \cite{li:GPU-accelerated-solvers,Naumov:Block-ILU-GPUs}, or sparse approximate inverses \cite{Dehnavi:SPAI-GPU,Lukash:SPAI,Sawyer:SPAI}.
Nevertheless, hardware-efficient and scalable black-box preconditioners for GPUs are not available, but instead the use of problem-specific information is required \cite{Yokota:Biomolecular-GPUs}.
Taking preconditioner setup costs into account, iterative solvers using simple diagonal preconditioners or no preconditioner at all are often observed to be competitive in terms of time-to-solution for small to mid-sized systems, where e.g.~the asymptotic optimality of multigrid preconditioners is not yet dominant \cite{Wagner:AMG}.
Similarly, matrix-free methods cannot be used with complicated black-box preconditioners in general.

In this work we consider three popular iterative solvers:
The conjugate gradient (CG) method for symmetric positive definite systems \cite{Stiefel:CG},
the stabilized bi-conjugate gradient (BiCGStab) method for non-symmetric positive definite systems \cite{vanderVorst:BiCGStab},
and the generalized minimum residual (GMRES) method for general systems \cite{saad-gmres}.
In contrast to previous work with a focus on the optimization of sparse matrix-vector products \cite{Ashari:SpMV-gpus,baskaran:sparse-matvec-ibm,Bell:SpMV-on-throughput-oriented-processors,Greathouse:CSR-adaptive,Kreutzer:Unified-sparse-matrix-format,Liu:Efficient-SpMV,Saule:SpMV-Xeon-Phi}, we consider the optimization potential of the full solvers rather than restricting the optimization to a single kernel.
After a careful evaluation of the limiting resources for different system sizes and different densities of nonzeros in the system matrix, pipelining and kernel fusion techniques are presented in Section~\ref{sec:techniques} to resolve these bottlenecks to the extent possible.
The key principle in pipelined techniques is to apply not only a single operation to a datum loaded from main memory, but to chain multiple operations together to reduce the overall number of loads and stores to global memory.
Pipelining is typically achieved by fusing multiple compute kernels, but compute kernels may also be fused only to reduce the overall number of kernel launches, not exhibiting any pipelining effect.
Pipelining and kernel fusion are then applied to the CG method, the BiCGStab method, and the GMRES method in Section~\ref{sec:solvers}, leading to more efficient solver implementations than those using a sequence of calls to the basic linear algebra subprograms (BLAS) in vendor-tuned libraries.
Section~\ref{sec:benchmark} then compares the proposed solver implementations with existing solver implementations for GPUs available in the software libraries CUSP\footnote{CUSP library: \url{http://cusplibrary.github.io/}}, MAGMA\footnote{MAGMA library: \url{http://icl.cs.utk.edu/magma/}}, and PARALUTION\footnote{PARALUTION library: \url{http://www.paralution.com/}}, demonstrating a substantial performance gain of pipelined implementations for small systems without sacrificing performance for large systems.
Our benchmark results demonstrate the benefit of kernel fusion and pipelining techniques for GPUs from AMD and NVIDIA for the CG method, the BiCGStab method, and the GMRES method, and clearly outline that these techniques have not been applied extensively in the context of GPU computing before.

The obtained execution times are also compared with those obtained from CPU-based implementations in the PETSc\footnote{PETSc library: \url{http://www.mcs.anl.gov/petsc/}} library to demonstrate that CPU-based implementations are superior for typical sparse systems below about $3\,000$ unknowns.
Our results, similar to previous investigations~\cite{Lee:Debunking-100x}, also falsify wide-spread misconceptions of extreme performance gains using GPUs.
We show that performance gains of GPUs over power-equivalent dual-socket CPU machines are  below an order of magnitude on average.
This holds true also for large problem sizes and when initial data setup costs on GPUs are not taken into account.
Finally, Section~\ref{sec:conclusion} discusses the implications of our findings to software design and the need for more tightly integrated future hardware generations.

%% file: techniques.tex
\section{Implementation Techniques for Fast Iterative Solvers} \label{sec:techniques}

\begin{figure}[tbp]
 \centering 
 \includegraphics[width=0.999\textwidth]{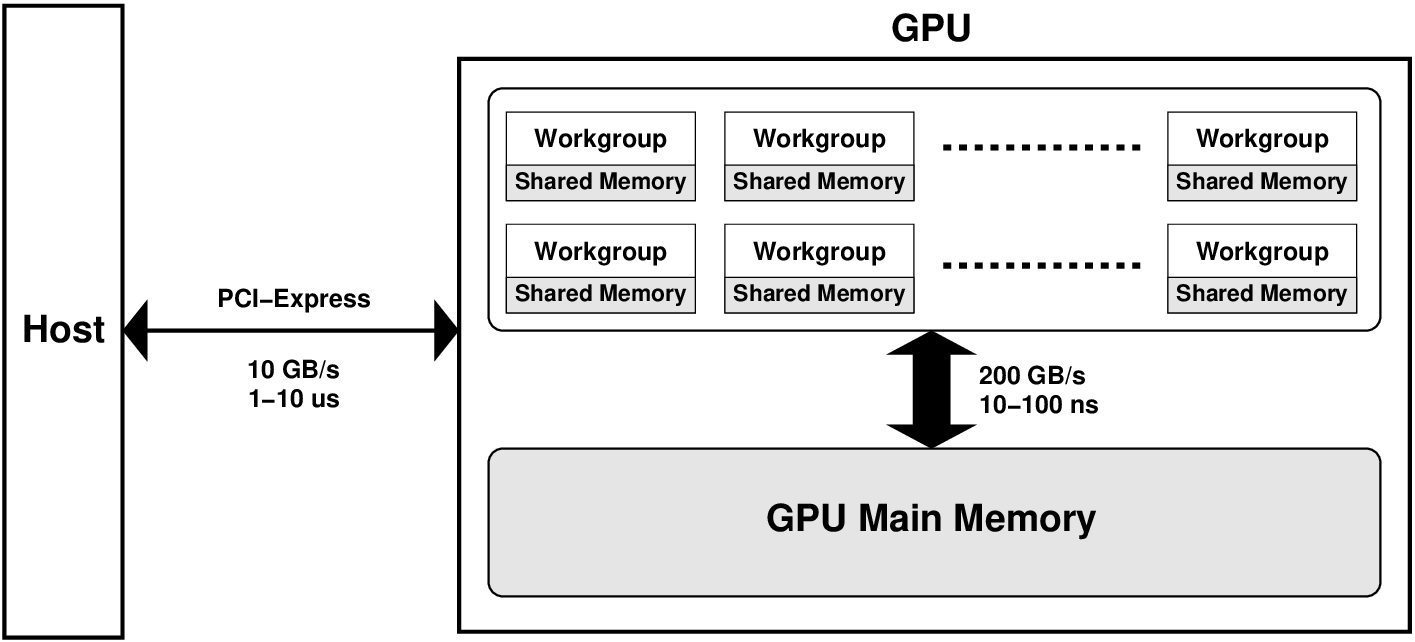} \\[0.1em]
 \caption{Schematic view of a GPU board connected to the host via PCI-Express at a bandwidth of about 10 GB/sec and a latency on the order of 10 microseconds.
          Each workgroup of threads can be synchronized through shared memory, but global synchronization is available only through separate kernel launches.
         }
 \label{fig:gpu-schematic}
\end{figure}

The purpose of this section is to identify the general bottlenecks of the typical building blocks of iterative solvers and to present techniques for mitigating their detrimental effects on performance.
A schematic view of a machine (\emph{host}) equipped with a discrete GPU connected via PCI-Express is given in Fig.~\ref{fig:gpu-schematic}, where the following key features are schematically depicted using a terminology similar to OpenCL:
\begin{itemize}
 \item Threads are collected in \emph{workgroups}, where each workgroup provides dedicated memory shared across threads in the workgroup.
       Thread synchronizations within a workgroup are possible inside a compute kernel, but a global synchronization of all workgroups is typically only possible by launching a new kernel.
       Although global synchronization primitives and spin locks through atomic operations are used occasionally, these techniques are not sufficiently portable across different hardware and thus not further considered.
 \item If a kernel launch is initiated on the host, it takes at least a few microseconds until the kernel will launch on a GPU.
       This is because a kernel launch on the GPU requires a message from the host to trigger the execution, entailing high latency for communication across PCI-Express.
       This latency of kernel launches can be hidden if another kernel is currently active on the GPU, in which case the PCI-Express message for launching the new kernel is received asynchronously.
 \item Memory access latency of GPU main memory is around three orders of magnitude smaller than the latency of messages across the PCI-Express bus.
 \item The memory bandwidth between GPU main memory and the GPU compute units can be more than ten times higher than the bandwidth of the PCI-Express bus connecting host and GPU.
       Current high-end GPUs offer over $200$ GB/sec memory bandwidth, whereas the current PCI-Express 3.0 offers up to $15.75$ GB/sec for a 16-lane slot.
\end{itemize}
The remainder of this section quantifies the overhead of PCI-Express latency and presents techniques for reducing the number of kernel launches to reduce the detrimental latency effect.

 \subsection{PCI-Express Latency}

\begin{figure}[tbp]
 \centering
 \includegraphics[width=0.48\textwidth]{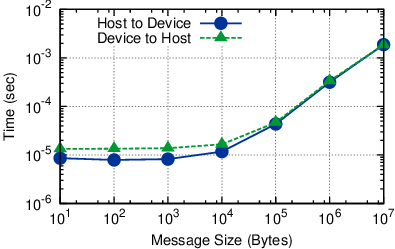} \hfill
 \includegraphics[width=0.48\textwidth]{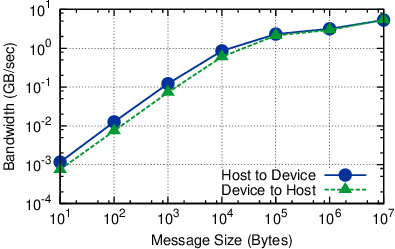}
 \caption{Plot of median values for execution time (left) and obtained bandwidth (right) from $100$ host-device data transfers over PCI-Express $2.0$ using an NVIDIA Tesla C2050.
          The benchmark uses the OpenCL functions \lstinline|clEnqueueWriteBuffer()| and \lstinline|clEnqueueReadBuffer()| in a blocking manner so that the respective function only returns after the data is sent or received.
          Message sizes below $10^4$ bytes are limited by latency, not PCI-Express bandwidth.}
 \label{fig:host-device}
\end{figure}

 At the very least, iterative solvers executed on the GPU need to communicate information about the current residual norm to the host.
 In the typical case of a communication of the residual norm in each iteration for convergence checks, the time required for a data transfer from the device to the host represents a lower bound for the time required for an iterative solver iteration.
 An OpenCL benchmark for PCI-Express data communication shown for an NVIDIA Tesla C2050 in Fig.~\ref{fig:host-device} exhibits a latency-dominated regime for message sizes below ten kilobytes, where the transfer time is around eight microseconds, and a bandwidth-limited regime for larger message sizes in accordance to the well-known idealized communication model based on latency and bandwidth~\cite{Foster:Designing-Building-Parallel-Programs}.
 Latency-dominated data transfer from the device to the host takes almost twice as long, because a transfer initiation from the host is required first.
 Similar timings and bandwidths are obtained on other GPUs both with PCI-Express 2.0 and 3.0.
 Our overall observation in Section~\ref{sec:benchmark} is that NVIDIA GPUs show slightly lower latency than AMD GPUs on the Linux-based machines used for the comparison.

 To better understand the latency induced by PCI-Express transfer, consider a high-end GPU with $200$ GB/s memory bandwidth.
 Within the PCI-Express latency of $8$ microseconds, the GPU can load or store $1.6$ megabytes of data from main memory assuming full saturation of the memory channel, which amounts to $200\,000$ values in double precision and which we will refer to as \emph{latency barrier}.
 Consequently, GPUs suffer from inherent performance constraints for any kernel limited by memory bandwidth whenever the total amount of data loaded or stored is significantly below the latency barrier.
 On the other hand, many practical applications induce systems with storage requirements for the unknowns close to or even below the latency barrier.
 In such case, iterative solver implementations for GPUs need to keep the latency-induced overhead as small as possible by packing multiple operations into each kernel.

  \subsection{Kernel Fusion}

As a prototypical example for many iterative solvers, consider the two operations
 \begin{align} \label{eq:fusion-example-1}
        q &= Ap \\
   \alpha &= \langle p, q \rangle \label{eq:fusion-example-2}
 \end{align}
 for a scalar value $\alpha$, vectors $p$ and $q$, and a sparse square matrix $A$.
 Conventional implementations based on BLAS routines involve the following steps:
 \begin{enumerate}
  \item Call the sparse matrix-vector product kernel for computing \eqref{eq:fusion-example-1}.
        For a standard compressed sparse row (CSR) representation of the sparse matrix, a typical OpenCL kernel body is as follows (cf.~\cite{baskaran:sparse-matvec-ibm,Bell:SpMV-on-throughput-oriented-processors}):
        \begin{lstlisting}
 for (uint i = get_global_id(0); i < size; i += get_global_size(0))
 {
   double q_at_i = 0;
   for (uint j = A_row[i]; j < A_row[i+1]; ++j)
     q_at_i += A_values[j] * p[A_col[j]];
   q[i] = q_at_i;
 }
        \end{lstlisting}
        where \lstinline|A_row| and \lstinline|A_col| are the arrays holding the row and column indices in the CSR storage format, and $\lstinline|A_values|$ holds the nonzero entries.
        High-performance implementations of sparse matrix-vector products for the CSR format are more involved than the simple example shown here \cite{Ashari:SpMV-gpus,Greathouse:CSR-adaptive}.

  \item Compute the partial results of $\langle p, q \rangle$ for the subvectors assigned to each of the thread workgroups.

  \item If $\alpha$ is stored on the GPU, launch another kernel using a single thread workgroup to sum the partial results.
        If $\alpha$ is stored on the host, transfer the partial results to the host, and perform the summation there.
 \end{enumerate}
 Although this conventional implementation can reuse vendor-tuned routines, the multiple kernel launches are detrimental to performance for data sizes below the PCI-Express latency barrier.

 On closer inspection, the operations \eqref{eq:fusion-example-1} and \eqref{eq:fusion-example-2} can be computed more efficiently by fusing compute kernels:
 Since the respective values in $q$ and $p$ are already available in the GPU processing elements when computing the matrix-vector product, they can be reused to compute the partial results for each thread workgroup of the inner product.
 The fused kernel body for the CSR format is as follows:
        \begin{lstlisting}
 //
 // Part 1: Matrix-vector product
 //
 double p_in_q = 0;
 for (uint i = get_global_id(0); i < size; i += get_global_size(0))
 {
   double q_at_i = 0;
   for (uint j = A_row[i]; j < A_row[i+1]; ++j)
     q_at_i += A_values[j] * p[A_col[j]];
   q[i]    = q_at_i;
   p_in_q += q_at_i * p[i];      // extra operation for <p, q>
 }

 //
 // Part 2: Reduction to obtain contribution from thread workgroups:
 //
 __local double shared_buf[BUFFER_SIZE];
 shared_buffer[get_local_id(0)] = p_in_q;
 for (uint stride=get_local_size(0)/2; stride > 0; stride /= 2)
 {
   barrier(CLK_LOCAL_MEM_FENCE);
   if (get_local_id(0) < stride)
     shared_buf[get_local_id(0)] += shared_buf[get_local_id(0) + stride];
 }

 if (get_local_id(0) == 0)
   partial_result[get_group_id(0)] = shared_buf[0];
        \end{lstlisting}
  First, the matrix-vector kernel from the previous snippet is only slightly augmented to accumulate the partial results for each thread in \lstinline|p_in_q|.
  Extra logic could be employed to explicitly avoid reading \lstinline|p[i]| from global memory if the respective diagonal entry of $A$ is nonzero, but \lstinline|p[i]| may still be available in cache anyway.
  Then, a reduction using a shared memory buffer \lstinline|shared_buf| of appropriate size \lstinline|BUFFER_SIZE| is applied to obtain the sum over all threads within a thread workgroup.
  Finally, the first thread in each thread workgroup writes the partial result of the workgroup to a temporary buffer \lstinline|partial_result|.
  The summation of the values in \lstinline|partial_result| is carried out on the host as outlined in the third step above.

\begin{figure}[tbp]
 \centering \vspace*{-0.2cm}
 \includegraphics[width=0.46\textwidth]{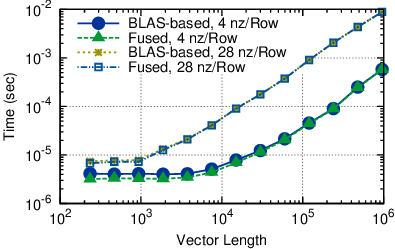} \hfill
 \includegraphics[width=0.46\textwidth]{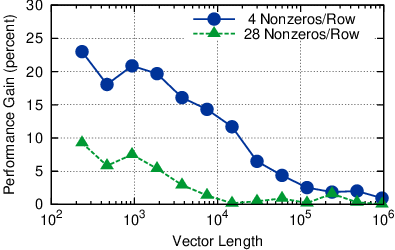} \\[-0.1em]
 \caption{Total time required to run the operations \eqref{eq:fusion-example-1} and \eqref{eq:fusion-example-2} for different matrix and vector sizes on an NVIDIA Tesla C2050.
          If the vector size is below $10\,000$ entries, the total time is dominated by the latency for enqueuing the kernel, not the kernel execution time.
         }
 \label{fig:kernel-fusion}
\end{figure}

  A comparison of execution times of the conventional implementation with the implementation using the fused kernel is given in Fig.~\ref{fig:kernel-fusion}.
  In both cases the final reduction step for the partial results from  $128$ thread workgroups has been computed on the host and is included in the timings.
  Two types of matrices have been compared:
  The first family of matrices with four randomly (with a uniform distribution over all column indices) distributed nonzeros per row is limited by latency for systems with up to $10^4$ unknowns.
  A performance gain of about $20$ percent is obtained from the use of a fused kernel, which reduces the number of kernels required from two to just one.
  At system sizes above $10^5$ unknowns, a performance gain of a few percent is still obtained because the vector $q$ does not need to be reloaded from memory when computing the inner product \eqref{eq:fusion-example-2}.
  The second matrix type with $28$ randomly distributed nonzeros per row is limited by the kernel execution time for system sizes below $10^3$ unknowns.
  This is because each thread needs to process $28$ nonzeros per row in $A$, which results in a larger execution time than the pure PCI-Express latency.
  Nevertheless, a performance gain of up to ten percent is obtained for smaller systems, yet there is no notable performance gain or loss at larger system sizes due to diminishing savings from reusing values from $q$ for computing the inner product.

  It is not only possible to fuse the first stage in the inner product computation with the matrix-vector product, but one can also fuse the second stage (summation of partial results) with subsequent operations.
  Since the summation result is usually needed in each thread workgroup, the final summation has to be computed in each thread workgroup in such case.
  These redundant computations are usually well below the PCI-Express latency barrier and thus faster than the use of a separate host-device transfer or a dedicated summation kernel.
  While kernel fusion can in principle be applied to an arbitrary number of vector updates, the global synchronization points induced by matrix-vector products as well as inner products are natural boundaries for fusing compute kernels.
  However, not every inner product induces a separate synchronization point: The partial summation stage of several inner products may also be computed within the same kernel, which is then followed by a second kernel computing the final results of the inner products and possibly other vector operations.

%% file: solvers.tex
\section{Pipelined Iterative Methods for Graphics Processing Units} \label{sec:solvers}
The implementation of the CG method, the BiCGStab method, and the GMRES method are investigated in depth in the following.
Each of these solvers is analyzed for the number of kernel launches to evaluate latency.
The kernel fusion techniques outlined in Section \ref{sec:techniques} are applied to reduce the number of kernel launches whenever appropriate.
We restrict our investigations to the execution on a single GPU, as this is the most frequent use case.
Nevertheless, the optimizations applied in this section can also be transferred to a multi-GPU setting, where additional data exchange between GPUs via the PCI-Express bus entails similar cost.
This allows, for example, to pack multiple partial results from inner products into a single memory buffer transfer to minimize latency.

 \subsection{Conjugate Gradient Method}
 Several variations of the classical CG method~\cite{Stiefel:CG} have been proposed in the past, cf.~\cite{Aliaga:Reformulated-CG,Barkai:Modified-CG,Chronopoulos:s-step-iterative-methods,Ghysels:Hiding-latency-cg}.
 Also, techniques for merging multiple solver iterations have been proposed, but they do not find broad acceptance in practice because of numerical instabilities \cite{Saad:Polynomial-preconditioning-cg}.
 In the following, the classical CG method and a pipelined version are compared, where the latter has already been developed for vector machines~\cite{Chronopoulos:s-step-iterative-methods}, revisited for extreme-scale scalability~\cite{Ghysels:Hiding-latency-cg}, and implemented in field-programmable gate arrays~\cite{Strzodka:Pipelined-mixed-precision-cg}:
\\[-1.em]
\begin{minipage}[t]{0.49\textwidth}
\null
 \begin{algorithm}[H]
    \caption{Classical CG} \label{alg:cg-classical}
    Choose $x_0$\;
    $p_0 = r_0 = b - Ax_0$\; \nllabel{alg:cg-classical-matvec0}
    \;
    \vspace*{0.1em}\;
    \vspace*{0.1em}\;
    \For{$i=0$ to convergence}{
      Compute and store $Ap_i$\; \nllabel{alg:cg-classical-matvec}
      Compute $\langle p_i, Ap_i\rangle$\; \nllabel{alg:cg-classical-dot1}
      $\alpha_i = \langle r_i, r_i\rangle / \langle p_i, Ap_i\rangle$\;
      $x_{i+1} = x_i + \alpha_i p_i$\;   \nllabel{alg:cg-classical-x}
      $r_{i+1} = r_i - \alpha_i A p_i$\; \nllabel{alg:cg-classical-r}
    \;
    \;
    \;
      Compute $\langle r_{i+1}, r_{i+1}\rangle$\;  \nllabel{alg:cg-classical-dot2}
    \vspace*{0.17em}\;
      $\beta_i = \langle r_{i+1}, r_{i+1}\rangle / \langle r_i, r_i\rangle$\;
      $p_{i+1} = r_{i+1} + \beta_i p_i$\; \nllabel{alg:cg-classical-p}
    }
  \end{algorithm}
\end{minipage}%
\begin{minipage}[t]{0.50\textwidth}
\null
 \begin{algorithm}[H] \label{alg:cg-pipelined}
    \caption{Pipelined CG}
    Choose $x_0$\;
    $p_0 = r_0 = b - Ax_0$\;   \nllabel{alg:cg-pipelined-matvec0}
    Compute and store $Ap_0$\; \nllabel{alg:cg-pipelined-matvec1}
    $\alpha_0 = \langle r_0, r_0\rangle / \langle p_0, Ap_0\rangle$\;
    $\beta_0 = \alpha_0^2 \langle Ap_0, Ap_0\rangle/\langle r_0, r_0\rangle - 1$\;
    \For{$i=1$ to convergence}{
    \;
    \;
    \;
      $x_i = x_{i-1} + \alpha_{i-1} p_{i-1}$\;   \nllabel{alg:cg-pipelined-x}
      $r_i = r_{i-1} - \alpha_{i-1} A p_{i-1}$\; \nllabel{alg:cg-pipelined-r}
      $p_i = r_i + \beta_{i-1} p_{i-1}$\;        \nllabel{alg:cg-pipelined-p}
      Compute and store $Ap_i$\;                 \nllabel{alg:cg-pipelined-matvec2}
      Compute $\langle Ap_i, Ap_i\rangle$, $\langle p_i, Ap_i\rangle$\; \nllabel{alg:cg-pipelined-dot1}
      Compute $\langle r_i, r_i\rangle$\;                               \nllabel{alg:cg-pipelined-dot2}
      $\alpha_i = (r_i, r_i\rangle / (p_i, Ap_i\rangle$\;                            \nllabel{alg:cg-pipelined-alpha}
      $\beta_i = \alpha_i^2 \langle Ap_i, Ap_i\rangle/\langle r_i, r_i\rangle - 1$\; \nllabel{alg:cg-pipelined-beta}
    \;
    }
  \end{algorithm}
\end{minipage}

  A direct implementation of Algorithm~\ref{alg:cg-classical} using one call to a matrix-vector product routine and five calls to BLAS routines per solver iteration is straightforward.
  Optimizations of the matrix-vector products on lines \ref{alg:cg-classical-matvec0} and \ref{alg:cg-classical-matvec} in Algorithm~\ref{alg:cg-classical} and lines \ref{alg:cg-pipelined-matvec0}, \ref{alg:cg-pipelined-matvec1}, and \ref{alg:cg-pipelined-matvec2} in Algorithm~\ref{alg:cg-pipelined} have been investigated in detail for different matrix formats on GPUs in the past~\cite{baskaran:sparse-matvec-ibm,Bell:SpMV-on-throughput-oriented-processors}.
  The inner products in lines \ref{alg:cg-classical-dot1} and \ref{alg:cg-classical-dot2} in the classical CG formulation impose synchronization by either splitting the operation into two kernels or by requiring a host-device transfer.
  In particular, the residual norm computed in line \ref{alg:cg-classical-dot2} is typically required on the host for convergence checks.
  The vectors $r$ and $p$ are loaded in lines \ref{alg:cg-classical-x} and \ref{alg:cg-classical-r}, but have to be reloaded for the search vector update operation in line \ref{alg:cg-classical-p}.

  The pipelined version in Algorithm \ref{alg:cg-pipelined} is based on the relation
  \begin{align} \label{eq:cg-recursion-beta}
   \beta_i = \frac{\langle r_{i+1}, r_{i+1}\rangle}{\langle r_i, r_i\rangle} = \frac{\alpha_i^2 \langle Ap_i, Ap_i\rangle - \langle r_i, r_i\rangle}{\langle r_i, r_i\rangle} = \alpha_i^2 \langle Ap_i, Ap_i\rangle/\langle r_i, r_i\rangle - 1
  \end{align}
  to compute $p_i$ in line \ref{alg:cg-classical-p} of Algorithm~\ref{alg:cg-classical} without having computed $\langle r_{i+1}, r_{i+1} \rangle$ yet.
  We note that it has been stated in the literature that precomputing inner products involving the vectors $p_i$ and $r_i$ by using recursion formulas based only on inner products of $p_j$ and $r_j$ with $j<i$ may lead to unstable algorithms \cite{Chronopoulos:s-step-iterative-methods,Saad:Polynomial-preconditioning-cg}.
  However, the computation of $\beta_i$ involves $Ap_i$, resulting in a stable algorithm based on experiences from multiple groups in different application contexts \cite{Barkai:Modified-CG,Chronopoulos:s-step-iterative-methods,Strzodka:Pipelined-mixed-precision-cg}.

  Relation \eqref{eq:cg-recursion-beta} allows for a rearrangement of Algorithm~\ref{alg:cg-classical} such that all vector updates can be computed right after each other (lines \ref{alg:cg-pipelined-x}, \ref{alg:cg-pipelined-r}, and \ref{alg:cg-pipelined-p} in Algorithm~\ref{alg:cg-pipelined}).
  An application of kernel fusion not only allows for computing all three vector updates within a single kernel instead of three, but also for avoiding a reload of $p_{i-1}$ and $r_i$ (line \ref{alg:cg-pipelined-p}) when using registers for intermediate values.
  Furthermore, all three inner products in Algorithm \ref{alg:cg-pipelined} can be computed simultaneously, allowing all intermediate results to be communicated to the host with a single data transfer.
  More precisely, the first reduction stage for the inner products $\langle Ap_i, Ap_i\rangle$, $\langle p_i, Ap_i\rangle$, and $\langle r_i, r_i\rangle$ can be computed within the same kernel sharing the same buffer for intermediate results.
  Then, the second reduction stage for obtaining the final results is either computed by only a single additional kernel launch, or by communicating all partial results with a single transfer to the host, which performs the summation.
  The data size of a single partial result is about one kilobyte, hence the data transfer time remains in the latency-dominated regime even if three of them are packed together, cf.~Fig.~\ref{fig:host-device}.

  To further enhance data reuse, we fuse the matrix-vector product in line \ref{alg:cg-pipelined-matvec2} with the inner products in line \ref{alg:cg-pipelined-dot1}, so that the result values of the matrix vector product can be processed right before they are written to GPU RAM.
  Thus, the computation of $\langle Ap_i, Ap_i\rangle$ and $\langle p_i, Ap_i\rangle$ comes at reduced data transfer cost, because the $j$-entry of $Ap_i$ has just been computed, and the $j$-th entry of $p_i$ may still be available in cache.
  Similarly, the inner product $\langle r_i, r_i\rangle$ in line \ref{alg:cg-pipelined-dot2} is fused with the vector update kernel for lines \ref{alg:cg-pipelined-x}, \ref{alg:cg-pipelined-r}, and \ref{alg:cg-pipelined-p}.

  In summary, we propose the following implementation of Algorithm \ref{alg:cg-pipelined}:
  \begin{itemize}
   \item Compute lines \ref{alg:cg-pipelined-x}, \ref{alg:cg-pipelined-r}, and \ref{alg:cg-pipelined-p} in one kernel and store the reduction results of each workgroup for the computation of $\langle r_i, r_i\rangle$ in line \ref{alg:cg-pipelined-dot2} in a temporary buffer.
   \item Compute lines \ref{alg:cg-pipelined-matvec2} and \ref{alg:cg-pipelined-dot1} in one kernel and append the reduction results of each workgroup for the computation of the inner products in line \ref{alg:cg-pipelined-dot1} to the same temporary buffer.
   \item Communicate the temporary buffer to the host, where the final reduction is computed to obtain $\langle r_i, r_i \rangle$, $\alpha_i$, and $\beta_i$ from lines \ref{alg:cg-pipelined-dot2}, \ref{alg:cg-pipelined-alpha}, and \ref{alg:cg-pipelined-beta}.
  \end{itemize}
  Since $\langle r_i, r_i\rangle$ is available for monitoring the residual norm on the host, a convergence check can be applied in each iteration with no extra effort.
  The proposed implementation requires only two kernel launches per iteration and one host-device data transfer.
  In contrast, a direct translation of the classical CG algorithm into BLAS-routines requires at least six kernel launches (lines \ref{alg:cg-classical-matvec}, \ref{alg:cg-classical-dot1}, \ref{alg:cg-classical-x}, \ref{alg:cg-classical-r}, \ref{alg:cg-classical-dot2}, and \ref{alg:cg-classical-p}) and may involve a second host-device data transfer for $\langle p_i, Ap_i\rangle$.
  Consequently, we expect an up to three-fold performance gain for small systems in the latency-dominated regime.
  Because $p_i$ and $r_i$ do not need to be loaded from memory twice per iteration, a performance gain of a few percent may also be obtained for large systems with very few nonzeros per row.

  If a fusion of the matrix-vector product in line \ref{alg:cg-pipelined-matvec2} and the partial reduction of the inner products in line \ref{alg:cg-pipelined-dot1} is not possible or desired, each of the two lines can be computed in separate kernels instead.
  This increases the number of kernel launches from two to three per iteration and requires one additional load and store operation of $Ap_i$ in global memory.
  Since the CUDA or OpenCL runtime can communicate all three kernel launches in a single transaction, no notable hit in the latency-dominated regime is expected.

 \subsection{BiCGStab}
 BiCGStab is an attractive iterative solver for systems described by nonsymmetric matrices, because the transposed operator $A^{\mathrm{T}}$ is not required.
 Based on the initial derivation \cite{vanderVorst:BiCGStab}, a pipelined method with only two global synchronizations has been proposed~\cite{Jacques:Modified-BiCGStab}.
 Later, a variant with only a single global synchronization has been proposed at the cost of an application of the transposed operator in the setup stage~\cite{Yang:Improved-BiCGStab}.
 Also, a preconditioned BiCGStab method overlapping global communication with the application of the preconditioner has been developed~\cite{Krasnopolsky:Reordered-BiCGStab}.
 A preliminary optimization study of the classical BiCGStab for GPUs is also available \cite{Anzt:BiCGStab-GPU}, for which we postpone a comparison to Section~\ref{sec:benchmark}.
 
 In analogy to the classical BiCGStab algorithm, the pipelined BiCGStab implementation proposed in this work does not require the transposed operator $A^{\mathrm{T}}$ to be available and is similar to the one proposed with two global synchronizations~\cite{Jacques:Modified-BiCGStab}.
 A comparison with the classical BiCGStab algorithm~\cite{saad-iterative-solution} is as follows:
\\[-1.em]
 \begin{minipage}[t]{0.49\textwidth}
\null
 \begin{algorithm}[H]
    \caption{Classical BiCGStab} \label{alg:bicgstab-classical}
    Choose $x_0$\;
    $p_0 = r_0 = b - Ax_0$\;
    Choose $r_0^*$ arbitrary\;
    Compute $\langle r_0, r_0^*\rangle$\;
    \For{$i=0$ to convergence}{
      Compute and store $Ap_i$\;                                            \nllabel{alg:bicgstab-classical-matvec1}
      Compute $\langle Ap_i, r_0^*\rangle$\;                                \nllabel{alg:bicgstab-classical-Ap-dot}
      $\alpha_i = \langle r_i, r_0^*\rangle / \langle Ap_i, r_0^*\rangle$\; \nllabel{alg:bicgstab-classical-alpha}
      $s_i      = r_i - \alpha_i Ap_i$\;                                    \nllabel{alg:bicgstab-classical-s}
      Compute and store $As_i$\;                                            \nllabel{alg:bicgstab-classical-matvec2}
      Compute $\langle As_i, s_i\rangle$, $\langle As_i, As_i\rangle$\;     \nllabel{alg:bicgstab-classical-As-dot}
      \;
      \vspace{1.5ex}\;
      $\omega_i = \langle As_i, s_i\rangle / \langle As_i, As_i\rangle$\;   \nllabel{alg:bicgstab-classical-omega}
      $x_{i+1}  = x_i + \alpha_i p_i + \omega_i s_i$\;                      \nllabel{alg:bicgstab-classical-x}
      $r_{i+1}  = s_i - \omega_i A s_i$\;                                   \nllabel{alg:bicgstab-classical-r}
      Compute $\langle r_{i+1}, r_0^*\rangle$\;                             \nllabel{alg:bicgstab-classical-r-dot}
      $\beta_i = \frac{\langle r_{i+1}, r_0^*\rangle}{\langle r_i, r_0^*\rangle} \times \frac{\alpha_i}{\omega_i}$\; \nllabel{alg:bicgstab-classical-beta}
      $p_{i+1} = r_{i+1} + \beta_i (p_i - \omega_i A_i)$\;                  \nllabel{alg:bicgstab-classical-p}
      \;
    }
  \end{algorithm}
\end{minipage}%
\begin{minipage}[t]{0.50\textwidth}
\null
 \begin{algorithm}[H]
    \caption{Pipelined BiCGStab} \label{alg:bicgstab-pipelined}
    Choose $x_0$\;
    $p_0 = r_0 = b - Ax_0$\;
    Choose $r_0^*$ arbitrary\;
    Compute $\langle r_0, r_0^*\rangle$\;
    \For{$i=0$ to convergence}{
      Compute and store $Ap_i$\;                                            \nllabel{alg:bicgstab-pipelined-matvec1}
      Compute $\langle Ap_i, r_0^*\rangle$\;                                \nllabel{alg:bicgstab-pipelined-Ap-dot}
      $\alpha_i = \langle r_i, r_0^*\rangle / \langle Ap_i, r_0^*\rangle$\; \nllabel{alg:bicgstab-pipelined-alpha}
      $s_i      = r_i - \alpha_i Ap_i$\;                                    \nllabel{alg:bicgstab-pipelined-s}
      Compute and store $As_i$\;                                            \nllabel{alg:bicgstab-pipelined-matvec2}
      Compute $\langle As_i, s_i\rangle$, $\langle As_i, As_i\rangle$\;     \nllabel{alg:bicgstab-pipelined-As-dots}
      Compute $\langle As_i, r_0^*\rangle$\;                                \nllabel{alg:bicgstab-pipelined-As-dot-r}
      $\beta_i = -\frac{(As_i, r_0^*\rangle}{\langle Ap_i, r_0^*\rangle}$\; \nllabel{alg:bicgstab-pipelined-beta}
      $\omega_i = \langle As_i, s_i\rangle / \langle As_i, As_i\rangle$\;   \nllabel{alg:bicgstab-pipelined-omega}
      $x_{i+1}  = x_i + \alpha_i p_i + \omega_i s_i$\;                      \nllabel{alg:bicgstab-pipelined-x}
      $r_{i+1}  = s_i - \omega_i A s_i$\;                                   \nllabel{alg:bicgstab-pipelined-r}
      \;
      \vspace{1.5ex}\;
      $p_{i+1} = r_{i+1} + \beta_i (p_i - \omega_i A_i)$\;                  \nllabel{alg:bicgstab-pipelined-p}
      Compute $\langle r_{i+1}, r_0^*\rangle$\;                             \nllabel{alg:bicgstab-pipelined-r-dot}
    }
  \end{algorithm}
\end{minipage}

 The classical BiCGStab method in Algorithm~\ref{alg:bicgstab-classical} requires a global synchronization after line~\ref{alg:bicgstab-classical-Ap-dot} to compute $\alpha_i$ for use in line~\ref{alg:bicgstab-classical-alpha}.
 Similarly, synchronizations are also required after line \ref{alg:bicgstab-classical-As-dot} to compute $\omega_i$ for use in line \ref{alg:bicgstab-classical-omega} and after line~\ref{alg:bicgstab-classical-r-dot} to compute $\beta$ for use in line~\ref{alg:bicgstab-classical-beta}.
 In analogy to the classical CG method, the search direction vector $p_{i+1}$ (line~\ref{alg:bicgstab-pipelined-p}) cannot be updated together with the approximated solution $x_{i+1}$ (line~\ref{alg:bicgstab-classical-x}) and the residual vector $r_{i+1}$ (line~\ref{alg:bicgstab-classical-r}).
 Consequently, additional loads from GPU main memory are required.
 Overall, two calls to routines for sparse matrix-vector products and at least eight calls to BLAS level 1 routines are needed in a conventional implementation of the classical BiCGStab method.
 Four host-device data transfers are required if each inner product induces a data transfer between host and device.
 An additional call to a BLAS level 1 routine and a host-device transfer are necessary if the residual norm is recomputed explicitly in each iteration.

 The pipelined BiCGStab version in Algorithm~\ref{alg:bicgstab-pipelined} allows for improved data reuse by shifting the calculation of $\beta_i$ to line \ref{alg:bicgstab-pipelined-beta} through
 \begin{align*}
  \beta_i &= \frac{\langle r_{i+1}, r_0^*\rangle}{\langle r_i, r_0^*\rangle} \times \frac{\alpha_i}{\omega_i} 
          = \frac{\langle s_i - \omega_i A s_i, r_0^*\rangle}{\langle r_i, r_0^*\rangle} \times \frac{\langle r_i, r_0^*\rangle}{\omega_i \langle Ap_i, r_0^*\rangle}
          = \frac{\langle s_i, r_0^*\rangle}{\omega_i \langle Ap_i, r_0^*\rangle} - \frac{\langle A s_i, r_0^*\rangle}{\langle Ap_i, r_0^*\rangle} \ .
 \end{align*}
 Using the orthogonality $\langle s_i, r_0^*\rangle = \langle r_i - \alpha_i Ap_i, r_0^*\rangle = 0 $ one arrives at
 \begin{align*}
   \beta_i = - \frac{\langle A s_i, r_0^*\rangle}{\langle Ap_i, r_0^*\rangle} \ ,
 \end{align*}
 which we found to be numerically stable based on our experiments.
 This derivation of a pipelined BiCGStab version is similar to the modification of the classical CG method in Algorithm~\ref{alg:cg-classical} to obtain the pipelined Algorithm~\ref{alg:cg-pipelined}.
 The minor price to pay for this rearrangement is the calculation of $\langle A s_i, r_0^*\rangle$ in line~\ref{alg:bicgstab-pipelined-As-dot-r}.

 The next step is to apply kernel fusion extensively to the pipelined BiCGStab version in Algorithm~\ref{alg:bicgstab-pipelined}.
 The calculation of $\langle A s_i, r_0^*\rangle$ can be fused with the sparse matrix product in line \ref{alg:bicgstab-pipelined-matvec2} together with the calculation of $\langle As_i, s_i\rangle$ and $\langle As_i, As_i\rangle$ in line \ref{alg:bicgstab-pipelined-As-dots}.
 Similarly, lines \ref{alg:bicgstab-pipelined-matvec1} and \ref{alg:bicgstab-pipelined-Ap-dot} are fused to a single kernel computing a matrix-vector product and the first reduction stage of the inner product.
 The vector update in line \ref{alg:bicgstab-pipelined-s} is fused with the second reduction stages for the inner products needed to compute $\alpha_i$ in line \ref{alg:bicgstab-pipelined-alpha}.
 Since the residual norm is obtained via
 \begin{align*}
  \langle r_{i+1}, r_{i+1} \rangle = \langle s_i, s_i \rangle - 2 \omega_i \langle s_i, As_i \rangle + \omega_i^2 \langle As_i, As_i \rangle
 \end{align*}
 for which $\langle s_i, As_i \rangle$ and $\langle As_i, As_i \rangle$ are computed in line \ref{alg:bicgstab-pipelined-As-dots} and needed for the calculation of $\omega_i$ in line \ref{alg:bicgstab-pipelined-omega}, we augment the update kernel for the computation of $s_i$ in line~\ref{alg:bicgstab-pipelined-s} with the first reduction stage for $\langle s_i, s_i \rangle$.
 The partial results are transferred to the host together with the partial results for all other inner products after line \ref{alg:bicgstab-pipelined-As-dot-r}, where $\beta_i$ and $\omega_i$ are computed.
 Finally, the vector updates in lines \ref{alg:bicgstab-pipelined-x}, \ref{alg:bicgstab-pipelined-r}, and \ref{alg:bicgstab-pipelined-p} as well as the first reduction stage for the inner product in line \ref{alg:bicgstab-pipelined-r-dot} are fused into another kernel.

 Overall, the proposed pipelined BiCGStab implementation of Algorithm \ref{alg:bicgstab-pipelined} consists of four kernel launches and one host-device transfer of the partial results from the four inner products $\langle As_i, s_i\rangle$, $\langle As_i, As_i\rangle$, $\langle As_i, r_0^*\rangle$, and $\langle s_i, s_i \rangle$:
 \begin{itemize}
  \item Compute the matrix-vector product in line \ref{alg:bicgstab-pipelined-matvec1} and the partial results for the two inner products required for $\alpha_i$ in line \ref{alg:bicgstab-pipelined-alpha}.
  \item Compute $s_i$ in line \ref{alg:bicgstab-pipelined-s} by redundantly computing $\alpha_i$ in each thread workgroup from the partial results of the inner products $\langle r_i, r_0^*\rangle$ and $\langle Ap_i, r_0^* \rangle$.
  \item Compute and store $As_i$ (line \ref{alg:bicgstab-pipelined-matvec2}) and the partial results for the inner products in lines \ref{alg:bicgstab-pipelined-As-dots} and \ref{alg:bicgstab-pipelined-As-dot-r}.
  \item Communicate all partial results for the inner products to the host, sum them there and perform a convergence check.
  \item Compute the vector updates in lines \ref{alg:bicgstab-pipelined-x}, \ref{alg:bicgstab-pipelined-r}, and \ref{alg:bicgstab-pipelined-p} as well as the partial results for the inner product in line \ref{alg:bicgstab-pipelined-r-dot}.
 \end{itemize}
 In comparison, the BiCGStab implementation proposed in \cite{Anzt:BiCGStab-GPU} requires five kernel launches and three reductions, while a BLAS-based implementation of the classical method requires at least eight kernel launches and four additional kernel launches or host-device transfers for the second reduction stage in the computation of the inner products.
 Therefore, a moderate improvement over the pipelined implementation in \cite{Anzt:BiCGStab-GPU} and a two- to three-fold performance gain over purely BLAS-based implementations in the latency-dominated regime are expected, assuming that kernel launches and host-device transfers entail comparable latency.

  If a fusion of the matrix-vector products in lines \ref{alg:bicgstab-pipelined-matvec1} and \ref{alg:bicgstab-pipelined-matvec2} with the partial reduction for the inner products in lines~\ref{alg:bicgstab-pipelined-alpha}, \ref{alg:bicgstab-pipelined-As-dots}, and \ref{alg:bicgstab-pipelined-As-dot-r} is not possible or desired, each of the two kernels can be split into one kernel for the matrix-vector product and one for the partial reductions.
  This increases the total number of kernel launches to six per iteration, of which the CUDA or OpenCL runtime can pack up to six kernel launches into a single communication, while preserving the benefit on only a single data transfer from the device to the host.

 \subsection{GMRES}
 In contrast to the CG and BiCGStab methods, the GMRES method requires to store the full Krylov basis rather than only the current search direction vector, leading to an increase in the number of operations with each iteration~\cite{saad-gmres}.
 To limit the computational expense, the GMRES method is typically restarted after $m$ iterations, which is denoted by GMRES($m$).
 Typical values for $m$ are in the range of $20$ to $50$.
 Smaller values tend to slow down the overall convergence, whereas higher values increase the computational cost and may lead to more time spent in the orthogonalization rather than the matrix-vector product, making GMRES less attractive when compared to other methods.

 Three methods for the computation of an orthonormal Krylov basis from a set of linearly independent vectors $\{v_k\}_{k=1}^m$ are common~\cite{saad-iterative-solution}, further algorithms employed for the orthogonalization in a multi-GPU setting with significantly different constraints in terms of communication can be found in~\cite{Yamazaki:CA-GMRES}:

 \begin{itemize}
  \item \emph{Classical Gram-Schmidt}: The $k$-th vector of the basis is obtained as
   \begin{align*}
     w_k \leftarrow \sum_{i=1}^{k-1} \langle v_i, v_k \rangle v_i \ ,
     v_k \leftarrow v_k - w_k
   \end{align*}
   followed by a normalization of $v_k$.
   The inner products $\langle v_i, v_k \rangle$ are independent and can be computed in parallel.
  \item \emph{Modified Gram-Schmidt}: An accumulation of round-off errors in the basis vectors $v_k$ may lead to a loss of orthogonality as the basis is augmented.
  Higher numerical robustness than for the classical Gram-Schmidt method has been observed for
   \begin{align*}
     v_k \leftarrow v_k - \langle v_i, v_k \rangle v_i
   \end{align*}
  for $i$ from $1$ to $k-1$ rather than forming a single update vector $w_k$.
  The disadvantage of the modified Gram-Schmidt method is the reduced parallelism: Instead of computing all inner products $\langle v_i, v_k \rangle$ concurrently, only one inner product can be computed at a time, followed by a vector update.
  \item \emph{Householder reflections}:
  The Krylov basis may also be obtained through Householder reflections $P_k = (I - \beta_k u_k u_k^{\mathrm{T}})$ with identity matrix $I$, suitably chosen scalars $\beta_k$, and Householder vectors $u_k$.
  Similar to the modified Gram-Schmidt method, the Householder reflections have to be applied sequentially to obtain the Krylov basis.
  Although the method allows for the computation of an orthonormal basis up to machine precision, the method is less regularly used for implementations of the GMRES method due to its sequential nature.
 \end{itemize}

 In the following we consider the simpler GMRES method \cite{walker-simpler-gmres}, which allows for a simpler solution of the minimization problem than the original formulation, but is otherwise comparable in terms of computational expense.
 A comparison of the restarted form and a pipelined formulation, both using the classical Gram-Schmidt method for higher efficiency on parallel architectures, is as follows:
\\[-1.em]
 \begin{minipage}[t]{0.51\textwidth}
\null
 \begin{algorithm}[H]
    \caption{Classical GMRES($m$)} \label{alg:gmres-classical}
    Choose $x_0$\;
    $r_0 = b - Ax_0$\;
    $\rho_0 = \Vert r_0 \Vert_2$\;
    $v_0 = r_0 = r_0 / \rho_0$\;
    $R_{i,j} = 0$ for $i,j \in \{1,\ldots, m\}$\;
    \For{$i=1$ to $m$}{
      $v_i = Av_{i-1}$\;
      \For{$j=1$ to $i-1$}{
       $R_{j, i} = \langle v_j, v_i \rangle$\;
      }
      \For{$j=1$ to $i-1$}{
        $v_i = v_i - R_{j, i} v_j$\;
      }
      $v_i = v_i / \Vert v_i \Vert$\;
      $\xi_i = \langle r, v_i \rangle$\;
      $r = r - \xi_i v_i$                 \nllabel{alg:gmres-classical-r}
    }
    \;
    \;
    \;
    Solve $R \eta = (\xi_1, \ldots, \xi_m)$\;
    Update $x_m = \eta_1 r + \sum_{i=2}^m \tilde{\eta_i} v_{i-1}$
  \end{algorithm}
\end{minipage}%
\hspace*{-0.3cm}
\begin{minipage}[t]{0.52\textwidth}
\null
 \begin{algorithm}[H]
    \caption{Pipelined GMRES($m$)} \label{alg:gmres-pipelined}
    Choose $x_0$\;
    $r_0 = b - Ax_0$\;
    $\rho_0 = \Vert r_0 \Vert_2$\;
    $v_0 = r_0 = r_0 / \rho_0$\;
    $R_{i,j} = 0$ for $i,j \in \{1,\ldots, m\}$\;
    \For{$i=1$ to $m$}{                                 \nllabel{alg:gmres-pipelined-for}
      $v_i = Av_{i-1}$\;                                \nllabel{alg:gmres-pipelined-matrix-vector}
      \For{$j=1$ to $i-1$}{
       $R_{j, i} = \langle v_j, v_i \rangle$\;          \nllabel{alg:gmres-pipelined-R}
      }
      \For{$j=1$ to $i-1$}{
        $v_i = v_i - R_{j, i} v_j$\;                    \nllabel{alg:gmres-pipelined-v-update}
      }
      $v_i = v_i / \Vert v_i \Vert$\;                   \nllabel{alg:gmres-pipelined-v-normalization}
      $\xi_i = \langle r, v_i \rangle$ (first stage)\;  \nllabel{alg:gmres-pipelined-xi}
    \;
    }
    \For{$i=1$ to $m$}{
      $\xi_i = \langle r, v_i \rangle$ (second stage)\;
    }
    Solve $R \eta = (\xi_1, \ldots, \xi_m)$\;             \nllabel{alg:gmres-solve-minimization}
    Update $x_m = \eta_1 r + \sum_{i=2}^m \eta_i v_{i-1}$ \nllabel{alg:gmres-pipelined-x}
  \end{algorithm}
\end{minipage}
 with $\tilde{\eta_i} = \eta_i + \eta_1 \xi_{i-1}$ to account for the updates of the residual $r$.

 The main difference between the classical formulation in Algorithm~\ref{alg:gmres-classical} and the pipelined formulation in Algorithm~\ref{alg:gmres-pipelined} involves the update of the residual vector in line~\ref{alg:gmres-classical-r} of Algorithm~\ref{alg:gmres-classical}.
 Because of the orthonormality of $\{v_k\}_{k=1}^m$, the inner product in line~\ref{alg:gmres-pipelined-xi} remains unchanged when using exact arithmetic.
 Similarly, since the values $\xi_i$ do not enter the Gram-Schmidt process, the values in the matrix $R$ remain unchanged so that round-off errors only affect the right hand side vector in line~\ref{alg:gmres-solve-minimization}.
 Our numerical experiments indicate that round-off errors in $\xi_i$ are dominated by round-off errors in the classical Gram-Schmidt process and therefore do not affect the overall numerical stability of the solver.
 Also, the convergence monitors proposed in \cite{walker-simpler-gmres} do not require updates of the residual and are based on the values $\xi_i$ only.
 Therefore, the full convergence history is still accessible before solving the minimization problem in line~\ref{alg:gmres-solve-minimization}, allowing for a correct handling of early convergence.
 Nevertheless, $m-1$ unnecessary steps of the Gram-Schmidt process will be carried out if convergence is obtained right at the first iteration, but this is rarely encountered for unpreconditioned solvers in practice.

 The benefit of removing the residual update from the Gram-Schmidt orthogonalization is that extensive kernel fusion can be applied to obtain an implementation of Algorithm~\ref{alg:gmres-pipelined} with almost no host-device communication.
 To begin, the reduction stage of the inner products in line~\ref{alg:gmres-pipelined-R} can be computed in two ways:
 The first option is a specialized matrix-vector routine for tall matrices if all Krylov vectors are stored as either the rows or the columns of a matrix.
 The second option is to fuse multiple inner products into the same kernel if all Krylov vectors reside in distinct buffers~\cite{Rupp:note-eigenvalues}.
 With both options the second reduction stage for computing $R_{j, i}$ in line \ref{alg:gmres-pipelined-R} is fused with the vector updates in line \ref{alg:gmres-pipelined-v-update} and also with the first reduction stage for computing $\Vert v_i \Vert$ needed in line \ref{alg:gmres-pipelined-v-normalization}.
 The normalization of $v_i$ in line \ref{alg:gmres-pipelined-v-normalization} is carried out by a kernel first computing the second reduction stage for $\Vert v_i \Vert$, then scaling $v_i$ and directly computing the first reduction stage for the obtaining $\xi_i$ in line \ref{alg:gmres-pipelined-xi}.
 Consequently, no data transfer between host and device is required during the Gram-Schmidt orthogonalization.
 An asynchronous transfer of the intermediate values for $\xi_i$ can be inserted at the end of each orthogonalization step for a better monitoring of the convergence process.

 After the Gram-Schmidt process, the intermediate results for computing $\xi_i$ are transferred to the host if not already transferred asynchronously, where the final values $\xi_i$ are computed.
 Similarly, the triangular matrix $R$ is transferred to the host.
 After the triangular system $R$ is inverted, the result vector containing the values $\eta_i$ is transferred to the device and the update of the result vector $x_m$ is computed in line \ref{alg:gmres-pipelined-x} using in a single kernel similar to the vector update in line \ref{alg:gmres-pipelined-v-update}.

 Overall, the proposed implementation of the pipelined GMRES($m$) method in Algorithm \ref{alg:gmres-pipelined} requires two kernel launches in the first iteration and four kernel launches in subsequent iterations:
 \begin{itemize}
  \item Compute the matrix-vector product in line~\ref{alg:gmres-pipelined-matrix-vector} and the first reduction stage for $\langle v_{i-1}, v_i \rangle$.
  \item Compute the first reduction stage for the inner products $\langle v_j, v_i \rangle$ in line \ref{alg:gmres-pipelined-R} with $j$ ranging from $1$ to $i-2$.
  \item Compute the second reduction stage for the inner products $\langle v_j, v_i \rangle$ in line \ref{alg:gmres-pipelined-R} for $j$ from $1$ to $i-1$, use the results directly for computing the vector update in \ref{alg:gmres-pipelined-v-update}, and compute the first reduction stage for $\Vert v_i \Vert$.
  \item Compute the second reduction stage for $\Vert v_i \Vert$, normalize $v_i$, and compute $\xi_i$.
 \end{itemize}
 A conventional implementation of the classical GMRES($m$) method in Algorithm~\ref{alg:gmres-classical} requires at least seven kernel launches and may involve several host-device data exchanges per iteration.
 Thus, an up to two-fold performance gain in the latency-dominated regime is expected.

 If a fusion of the matrix-vector product in line \ref{alg:gmres-pipelined-matrix-vector} and the first reduction stage for $\langle v_{i-1}, v_i \rangle$ is not possible or desired, each of the two operations can be computed in separate kernels instead.
 This increases the number of kernel launches from four to five per iteration and requires one additional load and store operation of $v_i$ in global memory.
 In light of the the subsequent inner products with the Krylov basis required for GMRES, these additional data loads and stores for $v_i$ are typically negligible.

 The computation of $\Vert v_i \Vert$ in Algorithm~\ref{alg:gmres-pipelined} can be avoided by making use of the shift-invariance property of the Krylov space, as it has been successfully demonstrated for $l^1$-GMRES in the context of large distributed-memory machines~\cite{Ghysels:Hiding-latency-GMRES}.
 This would allow for a reduction of the number of kernels from four to three, but the resulting GMRES variant would require an additional shift parameter.
 Moreover, since the Gram-Schmidt orthogonalization in Algorithm~\ref{alg:gmres-pipelined} is already free of intermediate host-device communication, the CUDA or OpenCL runtime can already communicate all kernel launches in a single PCI-Express message, thus no more gains from a further reduction of kernel launches are obtained.

 It is also worth comparing the pipelined $p^1$-GMRES method~\cite{Ghysels:Hiding-latency-GMRES} with Algorithm~\ref{alg:gmres-pipelined}.
 The former is concerned with overlapping global reductions in inner products with the computation of the sparse matrix-vector product involving local point-to-point communication on distributed memory machines.
 Such an overlap, however, is not needed in Algorithm~\ref{alg:gmres-pipelined}, because the whole orthogonalization phase is free from synchronizations with the host.

%% file: benchmarks.tex
\section{Benchmark Results} \label{sec:benchmark}
The implementations proposed in this work are implemented in the 1.7.0 release of the free open-source linear algebra library ViennaCL\footnote{ViennaCL library: \url{http://viennacl.sourceforge.net/}} and are compared in the following with the implementations in the free open-source libraries CUSP\footnote{CUSP library: \url{http://cusplibrary.github.io/}} 0.5.1, MAGMA\footnote{MAGMA library: \url{http://icl.cs.utk.edu/magma/}} 1.6.2 (linked with INTEL MKL 11.0), and PARALUTION\footnote{PARALUTION library: \url{http://www.paralution.com/}} 1.0.0.
Since CUSP and MAGMA are based on CUDA, benchmark data for AMD GPUs could only be obtained with ViennaCL and PARALUTION.
All four libraries are used in an out-of-the-box manner without additional target-specific tuning in order to reflect typical use cases.

In addition to classical implementations of the CG, BiCGStab, and GMRES methods, MAGMA also provides pipelined implementations of the CG method (four kernels with custom sparse matrix-vector product) and the BiCGStab method (nine kernels using vendor-tuned sparse matrix-vector products) \cite{Anzt:BiCGStab-GPU}.
Since MAGMA 1.6.2 provides a flexible GMRES implementation, but no classical GMRES implementation, we used the classical GMRES implementation in MAGMA 1.6.1 for comparison.
CUSP and PARALUTION implement classical formulations of all three iterative solvers in the comparison.
Our numerical experiments showed that the implementations of the classical CG and BiCGStab methods in MAGMA show similar performance to PARALUTION.
Therefore, we do not include timings for the classical implementations in MAGMA in our benchmark result plots, but instead only report execution times for the pipelined variants.

All tests were carried out on Linux-based machines running the CUDA 6.0 SDK on NVIDIA GPUs with GPU driver version 331.20 and the AMD APP SDK 2.9 with GPU driver version 13.352.1014 on AMD GPUs.
An NVIDIA Tesla C2050, an NVIDIA Tesla K20m, an AMD FirePro W9000, and an AMD FirePro W9100 were used for a comparison, representing the latest two generations of high-end workstation models from each vendor.
Error-correcting code memory was disabled on all four GPUs for better comparison.
Since all operations are limited by the available memory bandwidth, the obtained results are also representative for a broader range of mid- to high-end consumer GPUs with comparable memory bandwidth.

In addition to GPU-benchmarks, we also compare with the execution times obtained with the CPU-based PETSc\footnote{PETSc library: \url{http://www.mcs.anl.gov/petsc/}} library, version 3.6.0, on a dual-socket system equipped with INTEL Xeon E5-2620 CPUs, where parallel execution is based on the Message Passing Interface (MPI)\footnote{Message Passing Interface Forum: \url{http://www.mpi-forum.org/}} using MPICH\footnote{MPICH library: \url{http://www.mpich.org/}}~3.1.
The fastest execution time from runs with $1$, $2$, $4$, and $8$ MPI ranks for each system size are taken for comparison.
However, it should be noted that a comparison with a CPU-based library needs to be interpreted with care, because our benchmarks only compare the time taken per solver iteration, not the time required for copying the data to the GPU or for obtaining the result vector.

Execution times per iterative solver iteration are computed from the median value of ten solver runs with a fixed number of $30$ iterations for each solver.
In our experiments we have not observed any significant differences in the number of solver iterations required for convergence of the classical implementation and the pipelined implementation, hence the execution time per solver iteration is a suitable metric for comparison.

\begin{figure}[tbp]
 \centering
 \subfloat[Tesla C2050, CG      ]{ \includegraphics[width=0.32\textwidth]{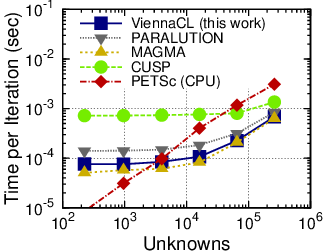} }
 \subfloat[Tesla C2050, BiCGStab]{ \includegraphics[width=0.32\textwidth]{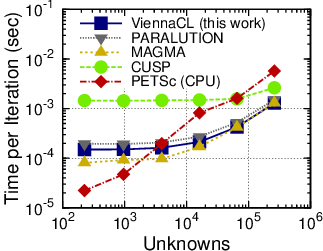} \label{fig:poisson2d-c2050bicgstab} }
 \subfloat[Tesla C2050, GMRES   ]{ \includegraphics[width=0.32\textwidth]{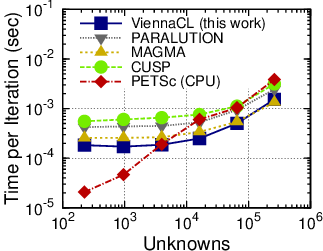} }

 \subfloat[Tesla K20m, CG      ]{ \includegraphics[width=0.32\textwidth]{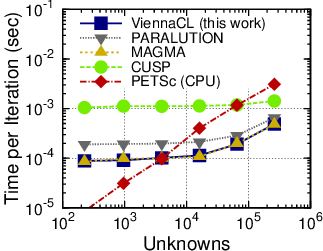} }
 \subfloat[Tesla K20m, BiCGStab]{ \includegraphics[width=0.32\textwidth]{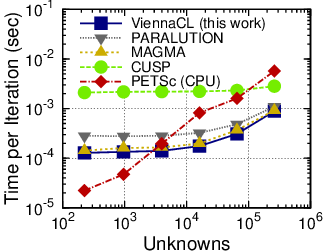} \label{fig:poisson2d-k20bicgstab}}
 \subfloat[Tesla K20m, GMRES   ]{ \includegraphics[width=0.32\textwidth]{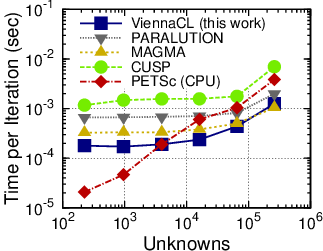} }

 \subfloat[FirePro W9000, CG      ]{ \includegraphics[width=0.32\textwidth]{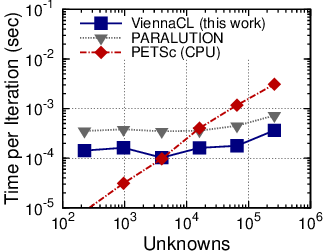} }
 \subfloat[FirePro W9000, BiCGStab]{ \includegraphics[width=0.32\textwidth]{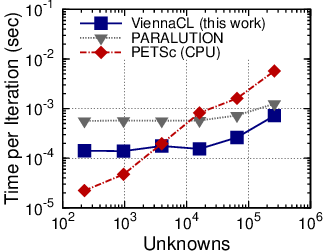} }
 \subfloat[FirePro W9000, GMRES   ]{ \includegraphics[width=0.32\textwidth]{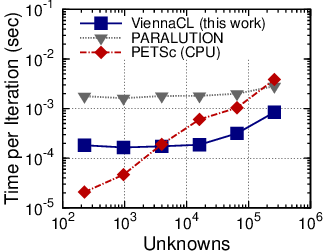} }

 \subfloat[FirePro W9100, CG      ]{ \includegraphics[width=0.32\textwidth]{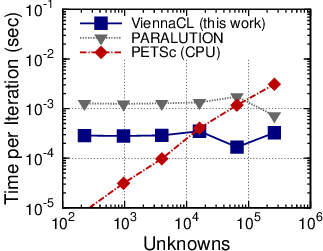} }
 \subfloat[FirePro W9100, BiCGStab]{ \includegraphics[width=0.32\textwidth]{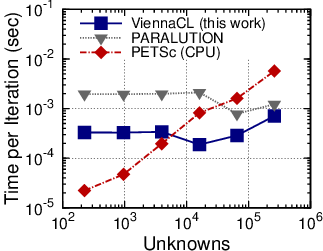} }
 \subfloat[FirePro W9100, GMRES   ]{ \includegraphics[width=0.32\textwidth]{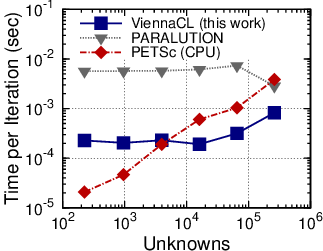} }

 \caption{Comparison of the solver time required per iteration for solving the Poisson equation using finite elements on triangular grids in two spatial dimensions.
          The proposed pipelined implementations in ViennaCL as well as the pipelined implementations in MAGMA outperform classical implementations in CUSP and PARALUTION for system sizes below $10^5$ thanks to a smaller number of kernel launches and better data reuse.}
 \label{fig:poisson2d}
\end{figure}

 \subsection{Linear Finite Elements for the Poisson Equation in 2D}
 We consider the execution time obtained with linear finite elements applied to the solution of the Poisson equation on the unit rectangle on a hierarchy of uniformly refined unstructured triangular meshes as a first benchmark.
 The resulting systems consist of $225$, $961$, $3\,969$, $16\,129$, $65\,025$, and $261\,121$ equations, respectively, and cover a broad range of typical system sizes solved on a single workstation.
 Results for CG, BiCGStab, and GMRES using the ELLPACK sparse matrix format (cf.~\cite{Bell:SpMV-on-throughput-oriented-processors} for a description) are given in Fig.~\ref{fig:poisson2d} for the four GPUs considered in our comparison.
 Similar results are obtained for other matrix formats, because the execution times in this setting are primarily dominated by latency effects.
 The case of large system matrices, where kernel execution times are dominant, is considered in Section~\ref{subsec:matrix-market}.

 Results for the NVIDIA GPUs show that the pipelined methods for the CG and the BiCGStab methods in ViennaCL and MAGMA show the same performance for small systems.
 The small differences can be explained by the use of asynchronous (and hence non-deterministic) convergence checks in MAGMA, whereas ViennaCL uses synchronous checks.
 Although MAGMA's BiCGStab implementation uses five kernels rather than the proposed implementation with four kernels, no significant difference is visible in Fig.~\ref{fig:poisson2d-c2050bicgstab} and Fig.~\ref{fig:poisson2d-k20bicgstab}.
 A comparison with the classical implementations in PARALUTION shows a roughly two-fold performance gain of pipelined implementations for small systems.
 The proposed pipelined GMRES implementation on NVIDIA GPUs is by about a factor of two faster in the latency-dominated regime than the implementation in MAGMA and about a factor of three faster than the implementations in CUSP and PARALUTION.
 
 On AMD GPUs the differences between ViennaCL and PARALUTION are more pronounced, because PARALUTION cannot take advantage from some optimizations in CUBLAS for NVIDIA GPUs.
 Conversely, these results suggests that the CUDA runtime for NVIDIA GPUs is able to hide the overhead of kernel launches more efficiently.
 A three-fold difference in execution times is obtained for the CG method, which reflects the different number of kernel launches, namely two for the pipelined implementation and six for a conventional implementation.
 A four-fold difference in execution times is obtained for the BiCGStab method, again reflecting the reduction in the number of kernel launches and reduced host-device communication in the proposed pipelined implementation.
 The difference for GMRES is approximately ten-fold, because the Gram-Schmidt orthogonalization in PARALUTION calls one kernel per dot-product during the orthogonalization procedure.
 
 Execution times for each solver iteration at system sizes below $10^4$ are practically constant for both NVIDIA and AMD GPUs.
 Because this constant is about a factor of two larger for AMD GPUs and because the AMD GPUs in this comparison offer higher memory bandwidth, essentially constant execution times are obtained for systems with up to $10^5$ unknowns for AMD GPUs.
 Only at system sizes above $10^5$ unknowns, PCI-Express communication becomes negligible compared to kernel execution times, hence the performance of all libraries becomes similar and varies only mildly.
 
 When comparing the execution times of GPU-based solvers with the execution times obtained with the CPU-based PETSc implementations, it is observed that the proposed pipelined implementations on GPUs are faster if systems carry more than about $3\,000$ unknowns on average.
 Depending on the underlying hardware and solver, up to $100\,000$ unknowns are needed with the conventional implementations in PARALUTION or CUSP to outperform the CPU-based implementations in PETSc.
 If initial data setup is taken into account, these cross-over points are shifted to even larger values, highlighting the importance of pipelining to increase the range of system sizes where GPU acceleration may pay off.

\begin{figure}[tbp]
 \centering
 \subfloat[Tesla C2050, CG      ]{ \includegraphics[width=0.32\textwidth]{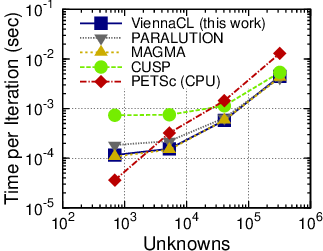} }
 \subfloat[Tesla C2050, BiCGStab]{ \includegraphics[width=0.32\textwidth]{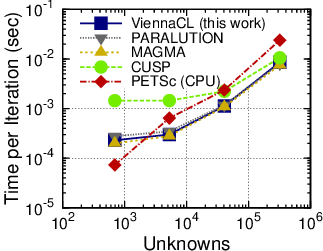} }
 \subfloat[Tesla C2050, GMRES   ]{ \includegraphics[width=0.32\textwidth]{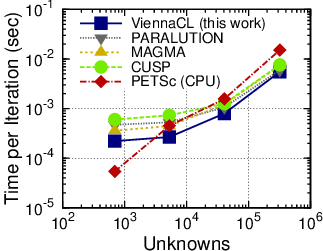} }

 \subfloat[Tesla K20m, CG      ]{ \includegraphics[width=0.32\textwidth]{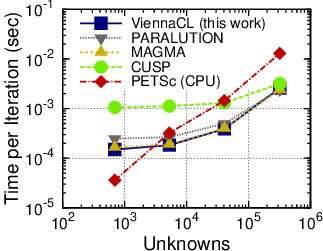} }
 \subfloat[Tesla K20m, BiCGStab]{ \includegraphics[width=0.32\textwidth]{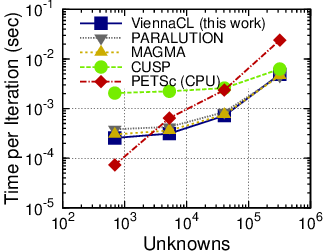} }
 \subfloat[Tesla K20m, GMRES   ]{ \includegraphics[width=0.32\textwidth]{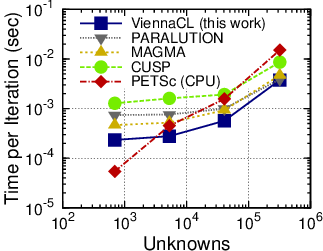} }

 \subfloat[FirePro W9000, CG      ]{ \includegraphics[width=0.32\textwidth]{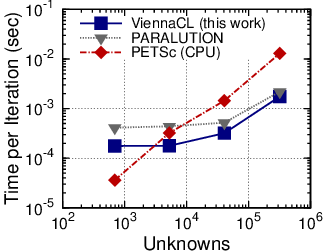} }
 \subfloat[FirePro W9000, BiCGStab]{ \includegraphics[width=0.32\textwidth]{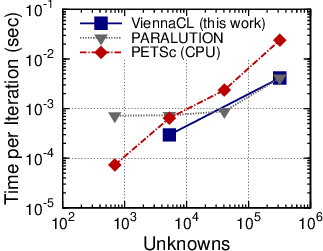} }
 \subfloat[FirePro W9000, GMRES   ]{ \includegraphics[width=0.32\textwidth]{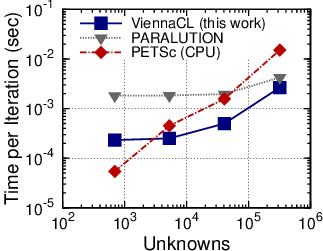} }

 \subfloat[FirePro W9100, CG      ]{ \includegraphics[width=0.32\textwidth]{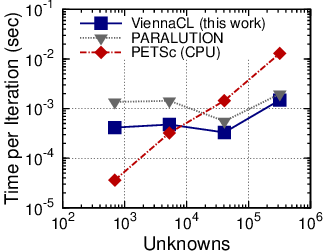} }
 \subfloat[FirePro W9100, BiCGStab]{ \includegraphics[width=0.32\textwidth]{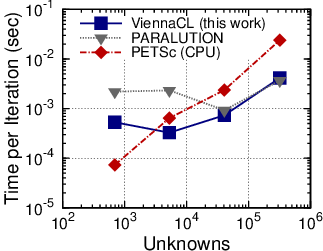} }
 \subfloat[FirePro W9100, GMRES   ]{ \includegraphics[width=0.32\textwidth]{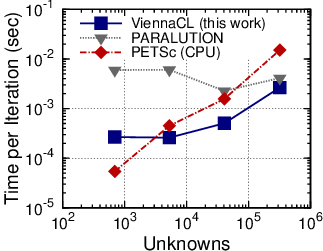} }

 \caption{Comparison of the solver time required per iteration for solving the linear elasticity model using finite elements in three spatial dimensions.
          The proposed pipelined implementations in ViennaCL as well as the pipelined implementations in MAGMA outperform other libraries for system sizes below $10^5$ thanks to a smaller number of kernel launches and better data reuse.}
 \label{fig:lame3d}
\end{figure}

 \subsection{Linear Finite Elements for Linear Elasticity in 3D}
 The second benchmark compares the execution time obtained with linear finite elements for numerical solutions of the linear elasticity model in three spatial dimensions.
 A hierarchy of uniformly refined tetrahedral meshes of the unit cube was used, resulting in system sizes of $693$, $5\,265$, $40\,725$, and $319\,725$, respectively.
 Compared to the first benchmark, the average number of unknowns per row increases from about $7$ to $60$ for the largest system, resulting in a higher share of the execution time being spent on sparse matrix-vector products.
 
 The results on NVIDIA GPUs in Fig.~\ref{fig:lame3d} show a similar trend as the results in Fig.~\ref{fig:poisson2d}:
 For small matrix sizes, the pipelined implementations of the CG and the BiCGStab methods in ViennaCL and MAGMA show similar performance.
 A two-fold performance gain over PARALUTION is obtained for the smallest system when using the CG method, which quickly diminish at larger system sizes due to more time spent on sparse matrix-vector products.
 While CUSP is about five times slower in the latency-limited regime for BiCGStab, the implementation in PARALUTION is less than a factor of two slower, suggesting that the CUDA-runtime is able to hide kernel launch latencies as well as host-device communication fairly well.
 Similar to the previous benchmark, the performance gain of the proposed pipelined implementation of GMRES is two-fold over MAGMA and three-fold over CUSP and PARALUTION.
 
 Performance differences between ViennaCL and PARALUTION on the AMD GPUs are about three-fold for the CG and BiCGStab methods. 
 For GMRES, a ten-fold performance advantage of the proposed pipelined implementation in the latency-dominated regime is obtained.
 
 Although the system matrix carries more nonzeros than in the first benchmark, about $2 \times 10^4$ unknowns on NVIDIA GPUs and $10^5$ unknowns on AMD GPUs are required such that kernel execution times hide performance penalties due to PCI-Express communication.

\subsection{Florida Sparse Matrix Collection} \label{subsec:matrix-market}

\begin{table}[tbp]
 \tbl{Summary of symmetric and non-symmetric matrices taken from the Florida Sparse Matrix Collection for comparison.
      These matrices represent the set of real-valued, floating point square matrices used in earlier contributions on optimizing sparse matrix-vector products \cite{Bell:SpMV-on-throughput-oriented-processors,Kreutzer:Unified-sparse-matrix-format}.
      \label{tab:matrix-summary}}{
  \begin{tabular}{|l|r|r|r|c|}
  \hline
  \textbf{Name}     & \textbf{Rows} & \textbf{Nonzeros} & \textbf{Nonzeros/Row} & \textbf{Symmetric} \\
  \hline \hline
  pdb1HYS           &     $36\,417$ &  $4\,344\,765$ & $119.31$    & yes \\
  cant              &     $62\,451$ &  $4\,007\,383$ &  $64.17$    & yes \\
  consph            &     $83\,334$ &  $6\,010\,480$ &  $72.13$    & yes \\
  shipsec1          &    $140\,874$ &  $7\,813\,404$ &  $55.46$    & yes \\
  pwtk              &    $217\,918$ & $11\,643\,424$ &  $53.39$    & yes \\
  \hline \hline
  rma10             &     $46\,835$ &  $2\,374\,001$ &  $50.69$    & no \\
  cop20k\_A         &    $121\,192$ &  $2\,624\,331$ &  $21.65$    & no \\
  scircuit          &    $170\,998$ &     $958\,936$ &   $5.61$    & no \\
  mac\_econ\_fwd500 &    $206\,500$ &  $1\,273\,389$ &   $6.17$    & no \\
  RM07R             &    $381\,689$ & $37\,464\,962$ &  $98.16$    & no \\
  Hamrle3           & $1\,447\,360$ &  $5\,514\,242$ &   $3.81$    & no \\
  kkt\_power        & $2\,063\,494$ & $13\,612\,663$ &   $7.08$    & no \\
  \hline
  \end{tabular}
 }
\end{table}

The performance of the proposed pipelined implementations is compared in the following for matrices from the Florida Sparse Matrix Collection\footnote{\url{http://www.cise.ufl.edu/research/sparse/matrices/}} used for the evaluation of sparse matrix-vector products in the past \cite{Bell:SpMV-on-throughput-oriented-processors,Kreutzer:Unified-sparse-matrix-format}.
While the focus in the previous section was on demonstrating the benefit of the proposed implementations for small to medium-sized systems, the purpose of this section is to show that the proposed implementations are also competitive for large systems.
Thus, the implementations in PARALUTION and MAGMA (for BiCGStab and GMRES) are a-priori expected to provide the best performance, since they use the vendor-tuned sparse matrix-vector product kernels from NVIDIA's CUSPARSE library.
In contrast, our implementations in ViennaCL rely on fused kernels, while CUSP implements the classical methods using its own set of sparse matrix-vector product kernels \cite{Bell:SpMV-on-throughput-oriented-processors}.

Since OpenCL does not support complex arithmetic natively, we restrict our benchmark to real-valued matrices listed in Tab.~\ref{tab:matrix-summary}.
The symmetric, positive definite matrices are used for benchmarking the implementations of the CG method, while the non-symmetric matrices are used for benchmarking the implementations of the BiCGStab and GMRES methods.
All sparse matrix formats available in the respective library are compared using implementations in CUDA and, if available, OpenCL.
The fastest combination is then taken for the comparison, since such a procedure resembles the typical user who picks the fastest sparse matrix format and the programming model with best performance for a particular application.


\begin{table}[tbp]
 \tbl{Comparison of the relative differences of residuals after $30$ solver iterations for the classical and the proposed pipelined algorithms.
      For CG and GMRES, the difference in residuals is only slightly above the inherent round-off error.
      The difference of the residuals obtained for the classical and the proposed pipelined BiCGStab method (Algorithm~\ref{alg:bicgstab-pipelined}) is larger, suggesting higher sensitivity with respect to round-off errors.
      \label{tab:residuals}}
  {
  \begin{tabular}{|l|c|c|c|}
  \hline
  \textbf{Matrix}   & \textbf{CG}           & \textbf{BiCGStab} & \textbf{GMRES} \\
  \hline \hline
  pdb1HYS           & $2.9 \times 10^{-12}$ & $1.9 \times 10^{-2}$ & $2.3 \times 10^{-15}$ \\
  cant              & $1.4 \times 10^{-14}$ & $1.2 \times 10^{-6}$ & $2.8 \times 10^{-11}$ \\
  consph            & $3.0 \times 10^{-15}$ & $7.3 \times 10^{-7}$ & $9.8 \times 10^{-10}$ \\
  shipsec1          & $7.4 \times 10^{-12}$ & $1.4 \times 10^{-2}$ & $4.0 \times 10^{-10}$ \\
  pwtk              & $3.0 \times 10^{-14}$ & $1.2 \times 10^{-6}$ & $6.5 \times 10^{-11}$ \\
  \hline \hline
  rma10             & -                     & $4.1 \times 10^{-1}$ &  $5.3 \times 10^{-8\hphantom{1}}$ \\
  cop20k\_A         & -                     & $3.4 \times 10^{-6}$ &  $1.8 \times 10^{-11}$ \\
  scircuit          & -                     & $1.4 \times 10^{-2}$ &  $2.1 \times 10^{-8\hphantom{1}}$ \\
  mac\_econ\_fwd500 & -                     & $1.5 \times 10^{-1}$ &  $4.6 \times 10^{-14}$ \\
  RM07R             & -                     & $2.2 \times 10^{-1}$ &  $1.4 \times 10^{-11}$ \\
  Hamrle3           & -                     & $1.1 \times 10^{-1}$ &  $1.1 \times 10^{-16}$ \\
  kkt\_power        & -                     & $4.7 \times 10^{-2}$ &  $4.9 \times 10^{-12}$ \\
  \hline
  \end{tabular}
 }
\end{table}

A comparison of the relative difference of the residuals obtained for the classical and the pipelined solvers after $30$ solver iterations is given in Table~\ref{tab:residuals}.
For CG and GMRES, the relative differences are below $10^{-10}$ for all matrices considered, hence the classical and the pipelined methods can be considered to be equally stable.
In contrast, the relative differences of the residuals obtained for BiCGStab are up to $41$ percent~(rma10), where differences are larger if BiCGStab converges slower or even stagnates.
This suggests that the classical BiCGStab method in Algorithm~\ref{alg:bicgstab-classical} and the pipelined BiCGStab method in Algorithm~\ref{alg:bicgstab-pipelined} show different sensitivity with respect to round-off errors.
However, as the relative differences remain below unity and as the residual norms for the pipelined BiCGStab method are smaller than those for the classical method for seven out of twelve matrices, we conclude that neither of the two methods is more sensitive to round-off errors than the other.


\begin{figure}[tbp]
 \centering
 \includegraphics[width=0.96\textwidth]{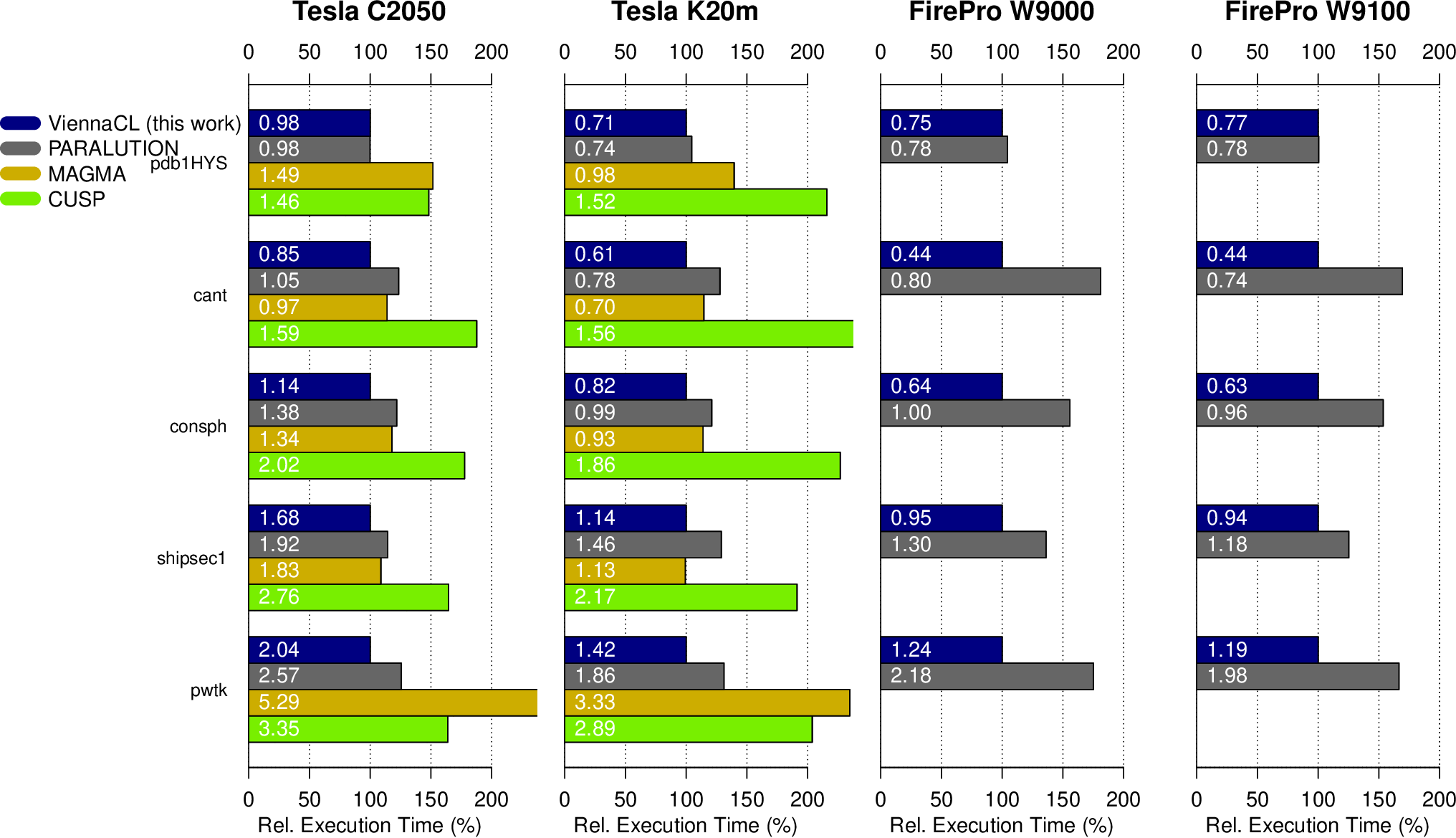}
 \caption{Comparison of execution times per CG solver iteration for different systems from the Florida Sparse Matrix Collection relative to the proposed pipelined implementations.
          Absolute execution times in milliseconds are given inside each bar. ViennaCL implements the pipelined methods proposed in this work, MAGMA uses a similar pipelined implementation without using vendor-tuned kernels.}
 \label{fig:mmarket-cg}
\end{figure}

The benchmark results for the CG method in Fig.~\ref{fig:mmarket-cg} show that the proposed solver implementation provides the best overall performance on all four devices.
Although the pipelined CG method implemented in MAGMA is similar to the one proposed here, the performance difference reflects the importance of providing fast fused kernels.
Similarly, the difference is particularly pronounced on AMD GPUs, where the performance of our proposed implementation is up to twice as high as the performance of PARALUTION, which needs to rely on its own kernels rather than using vendor-tuned implementations.
A comparison of absolute execution times also shows that the AMD GPUs provide a better overall performance due to their higher memory bandwidth.


\begin{figure}[tbp]
 \centering
 \includegraphics[width=0.96\textwidth]{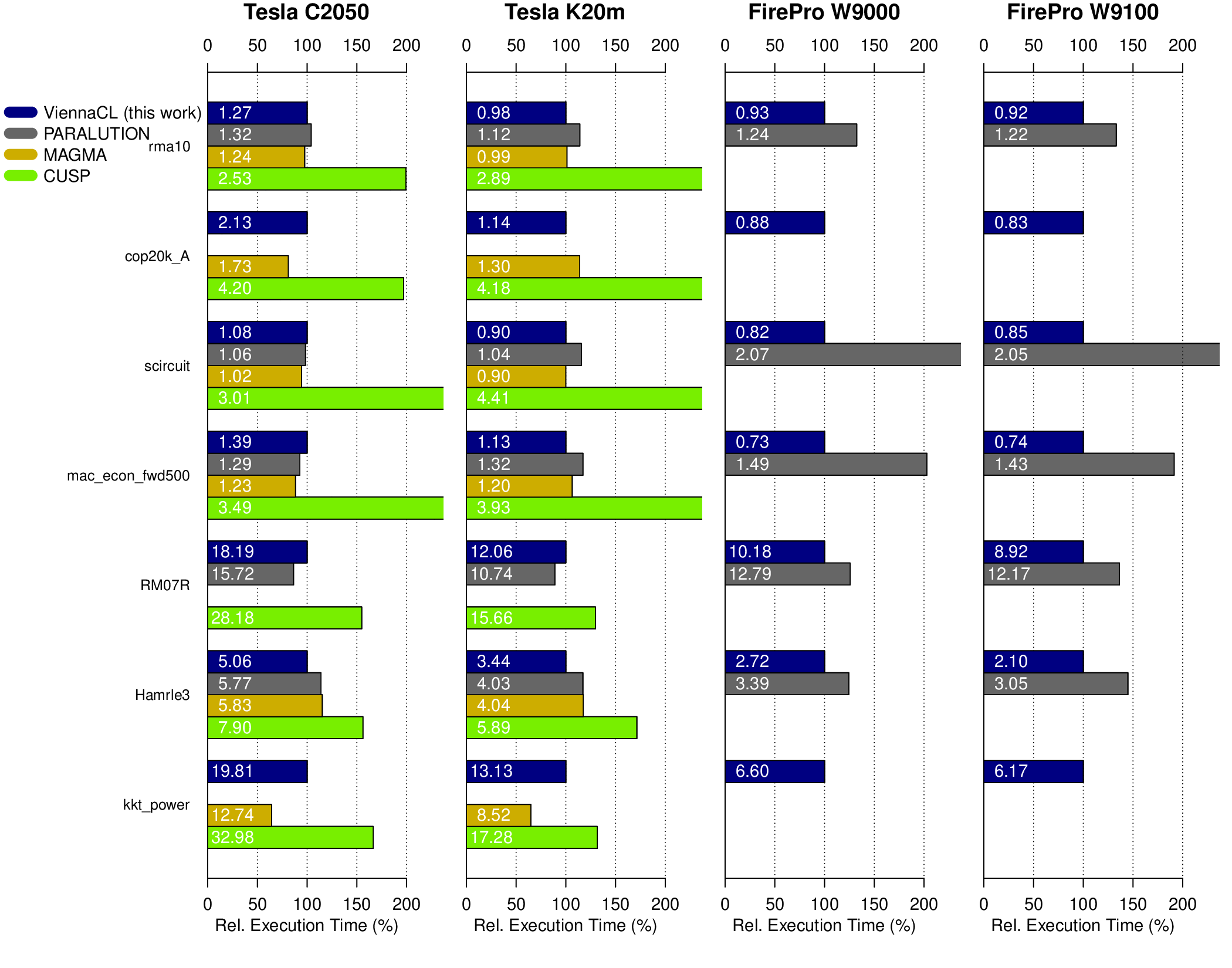}
 \caption{Comparison of execution times per BiCGStab solver iteration for different systems from the Florida Sparse Matrix Collection relative to the proposed pipelined implementations.
          Absolute execution times in milliseconds are given inside each bar.
          The \texttt{cop20k\_A} and the \texttt{kkt\_power} matrices could not be tested with PARALUTION due to segmentation faults.
          The \texttt{RM07R} matrix could not be run with MAGMA since it did not pass a check for positive definiteness.}
 \label{fig:mmarket-bicgstab}
\end{figure}

The comparison of execution times for the BiCGStab method in Fig.~\ref{fig:mmarket-bicgstab} shows similar performance of ViennaCL, PARALUTION, and MAGMA for NVIDIA GPUs on average:
Depending on the device and the matrix considered, either of the three is the best choice.
Since the proposed implementations do not contain any device-specific or matrix-specific optimizations, further tuning may provide further performance gains.
In contrast, the use of vendor-tuned kernels for the implementations in PARALUTION and MAGMA imposes limitations on further device- or matrix-specific tweaks to what is offered by the vendor library.
The custom sparse matrix-vector product kernels in CUSP result in about 60 percent higher execution times on average.
On AMD GPUs, the performance gain over PARALUTION is about 50 percent on average.
Similar to the results of the benchmark of the CG method, slightly higher overall performance can be obtained on AMD GPUs because of their higher memory bandwidth.


\begin{figure}[tbp]
 \centering
 \includegraphics[width=0.96\textwidth]{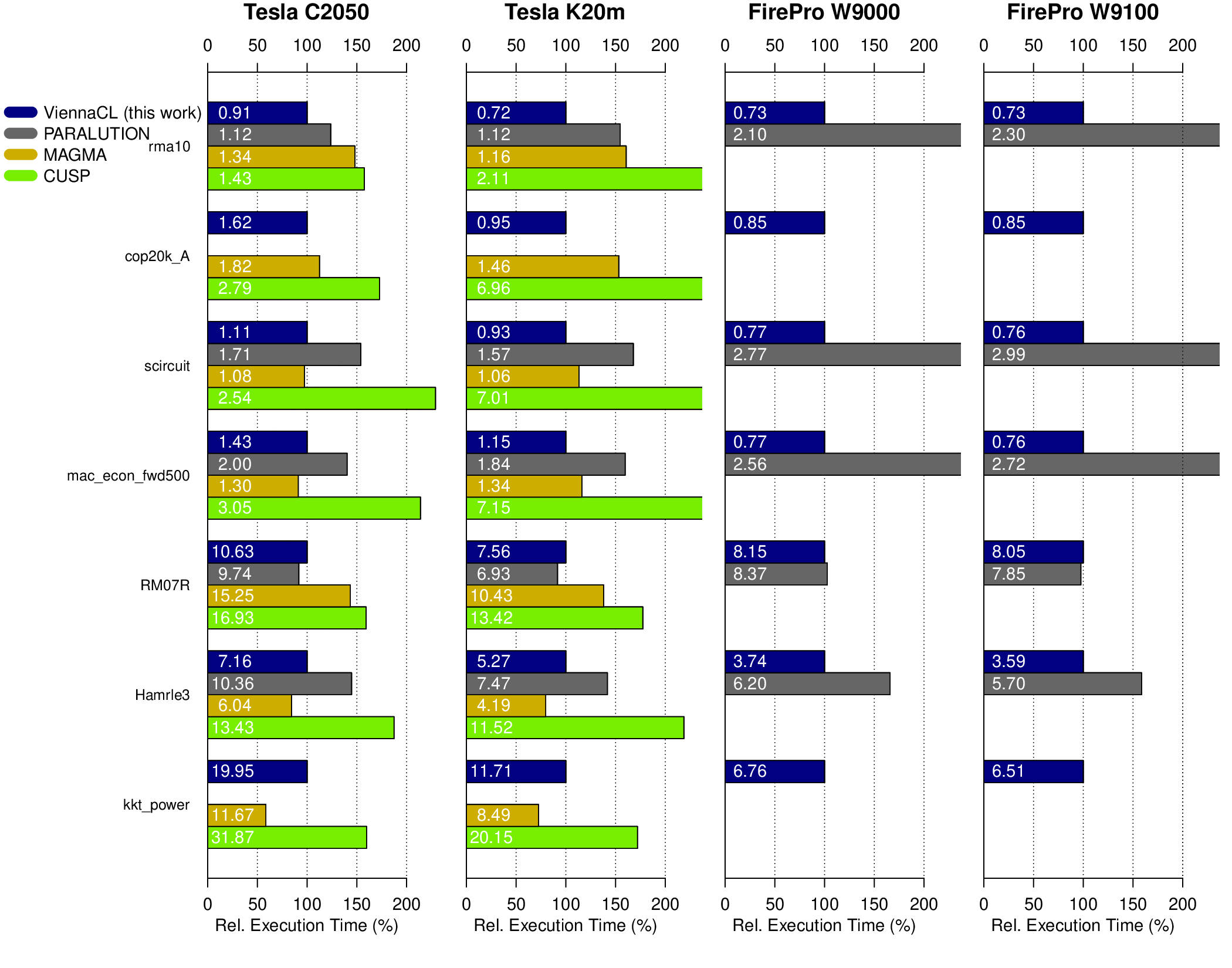}
 \caption{Comparison of execution times per GMRES solver iteration for different systems from the Florida Sparse Matrix Collection relative to the proposed pipelined implementations.
          Absolute execution times in milliseconds are given inside each bar.
          The \texttt{cop20k\_A} and the \texttt{kkt\_power} matrices could not be tested with PARALUTION due to segmentation faults.
         }
 \label{fig:mmarket-gmres}
\end{figure}

The benchmark results obtained for the GMRES method are depicted in Fig.~\ref{fig:mmarket-gmres} and show the same trend as the results obtained when comparing the implementations of the BiCGStab method.
Depending on the device and the matrix considered, either of ViennaCL, PARALUTION, and MAGMA may be the best choice.
In particular, no performance penalty from using pipelined implementations for large systems can be observed.


\begin{table}[tbp]
 \tbl{Relative share of the execution time per solver iteration spent on the sparse matrix-vector product, evaluated on an NVIDIA Tesla K20m.
      While the execution time for the CG and the BiCGStab method are usually dominated by computing sparse matrix-vector products particularly after pipelining, orthogonalizations in the GMRES method dominate.
      \label{tab:spmv}}
  {
  \begin{tabular}{|l|c||l|c|c|}
  \hline
  \textbf{Matrix}   & \textbf{CG}  & \textbf{Matrix}   &  \textbf{BiCGStab} & \textbf{GMRES} \\
  \hline \hline
  pdb1HYS           & $79.9 \%$    & rma10             & $78.2 \%$ & $53.2 \%$ \\
  cant              & $89.5 \%$    & cop20k\_A         & $89.3 \%$ & $53.6 \%$ \\
  consph            & $89.0 \%$    & scircuit          & $44.0 \%$ & $21.3 \%$ \\
  shipsec1          & $89.2 \%$    & mac\_econ\_fwd500 & $50.1 \%$ & $24.6 \%$ \\
  pwtk              & $88.9 \%$    & RM07R             & $91.2 \%$ & $72.8 \%$ \\
                    &              & Hamrle3           & $52.3 \%$ & $17.1 \%$ \\
                    &              & kkt\_power        & $58.9 \%$ & $32.9 \%$ \\
  \hline
  \end{tabular}
 }
\end{table}

The relative share of execution time spent on just computing matrix vector products by running isolated sparse matrix-vector product kernels as compared to full solver cycles is given in Table~\ref{tab:spmv}.
About $85$ percent of the time is spent on matrix-vector products for the CG method after pipelining, so significant reductions in execution times can only be obtained by optimizing the sparse matrix-vector product.
Similarly, $66$ percent of the time is spent on matrix-vector products in the pipelined BiCGStab method on average, where the share correlates well with the average number of nonzeros per row.
For the GMRES method, however, $60$ percent of the time is spent outside the matrix-vector product on average, justifying the careful optimization of the orthogonalization of the Krylov vectors via kernel fusion and pipelining.


\begin{figure}[tbp]
 \centering
 \includegraphics[width=0.99\textwidth]{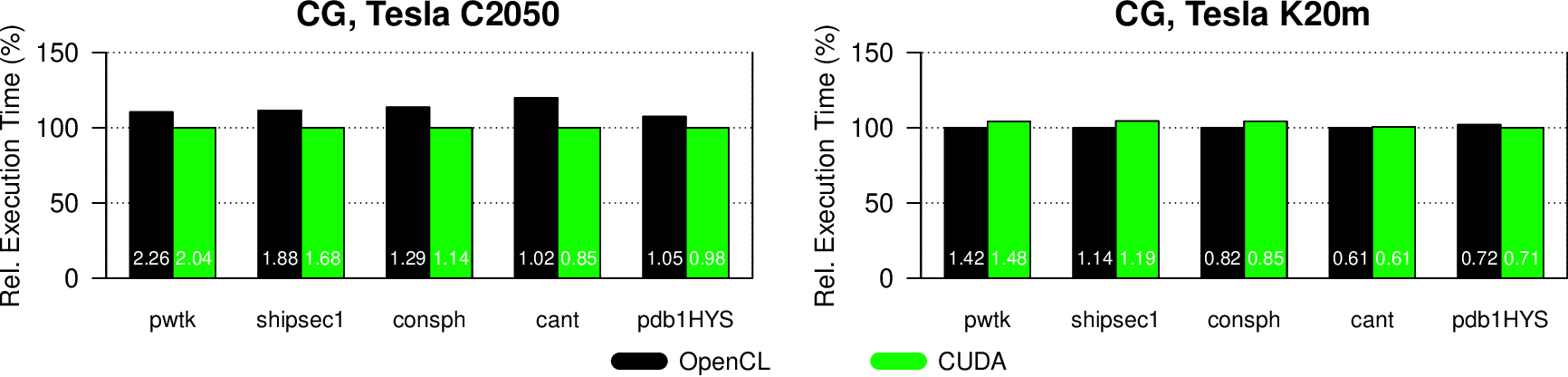} \\[0.5em]
 \includegraphics[width=0.475\textwidth]{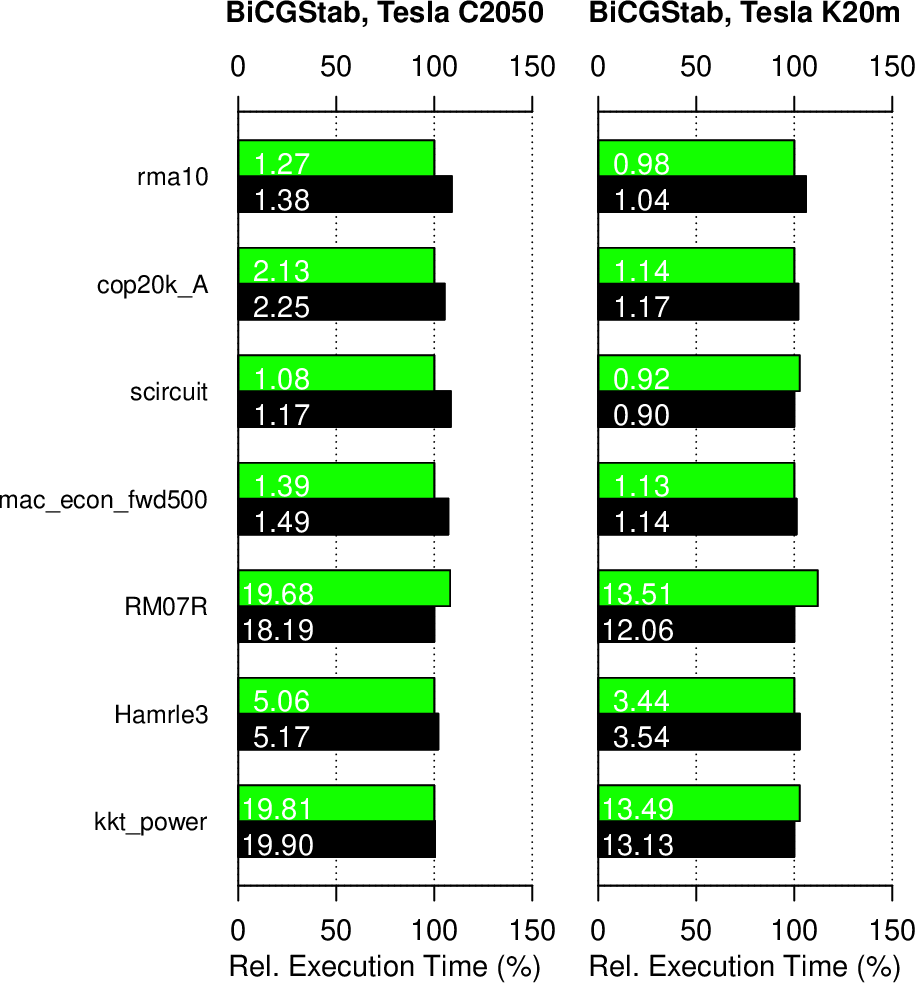} \hfill
 \includegraphics[width=0.47\textwidth]{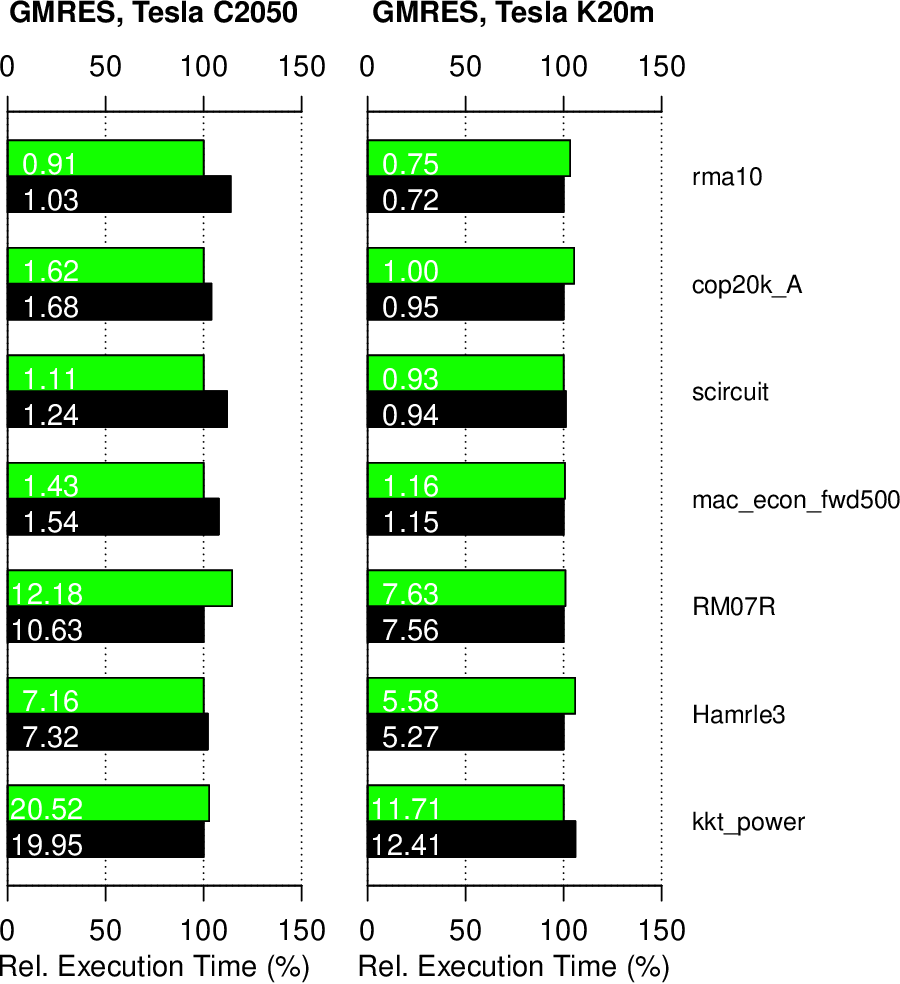}
 \caption{Comparison of execution times obtained with CUDA and OpenCL for the CG method (top), the BiCGStab method (left), and the GMRES method (right).
          Relative execution times are with respect to the faster framework.
          Absolute execution times in milliseconds are given inside each bar.
          Overall, the performance differences of CUDA and OpenCL are negligible in practice, even though OpenCL shows slightly better performance overall. }
 \label{fig:mmarket-cuda-opencl}
\end{figure}

Finally, execution times for the proposed implementations of the three iterative solvers using CUDA and OpenCL are compared in Fig.~\ref{fig:mmarket-cuda-opencl}.
In all cases, the obtained execution times of CUDA and OpenCL are within a few percent, which is a negligible difference in practice.

Overall, the benchmark results confirm that pipelined methods are not only favorable for smaller systems, where latency effects are significant, but also competitive for large systems.
This is beneficial for code maintenance, as only a single implementation needs to be maintained.
Furthermore, our results also suggest that a single implementation in OpenCL is sufficient, as the performance differences to CUDA are negligible.
This, however, faces practical limitations, as user codes may be written only in CUDA and thus incompatible with OpenCL.

%% file: conclusion.tex
\section{Conclusion} \label{sec:conclusion}
The proposed pipelined implementations of the CG, BiCGStab, and GMRES methods address the latency-induced performance penalties of GPU-accelerated implementations for sparse systems with less than about $10^5$ unknowns.
Our comparison with other solver packages shows significant performance gains over conventional implementations for practically relevant problem sizes between $10^4$ and $10^5$ unknowns.
A comparison for larger systems shows that the proposed implementations using fused kernels provide a performance competitive with implementations built on top of vendor-tuned kernels.
As a consequence, our results suggest that future efforts on the optimization of compute kernels should not be restricted to standard BLAS or BLAS-like kernels, but additional performance can be obtained if also optimized implementations for fused kernels are provided.
For example, not only the sparse matrix-vector product kernel, but also a kernel computing the sparse matrix-vector product plus the first reduction stage of inner products involving the result vector may offer superior performance for iterative solvers from the family of Krylov methods.

While an extensive use of pipelining and kernel fusion addresses latency issues and limited memory bandwidth, it also brings new challenges for the design of scientific software.
To leverage the full potential of modern hardware, it is no longer sufficient to only use a fairly small set of vendor-tuned BLAS-kernels, but instead provide modular building blocks for minimizing communication of data.

Future GPUs as well as CPUs will see gains in memory bandwidth, but the latency induced by the PCI-Express bus will not change substantially.
Therefore, the minimum system size required to get any performance gains on GPUs over CPUs will continue to grow.
As a consequence, the replacement of the PCI-Express bus with a interconnect technology of lower latency is essential for making accelerators more attractive.
Integrations of GPU units on the CPU die are one possible path to achieve lower latency.
However, no benefit over a well-optimized, purely CPU-based implementations can be expected for the memory-bandwidth limited operations in iterative solvers, if both the accelerator and the CPU core share the same memory link.

The techniques applied in this work can also be extended to preconditioned iterative solvers.
Not only can the application of the preconditioner be possibly fused with vector updates, but also the setup stage can benefit from fusing as many operations as possible into the same kernel.
A rigorous application of these techniques to preconditioners is left for future work.